\documentclass[aps,prd]{revtex4-1}
\usepackage{color,graphicx}
\usepackage{subfig}
\usepackage{ifpdf}
\usepackage{amsmath}
\usepackage[english]{babel}
\usepackage{bm}
\usepackage{color}
\usepackage{cancel}
\usepackage[T1]{fontenc}
\usepackage[utf8]{inputenc}

\begin{document}

\title{Restrained Dark $U(1)_d$ at Low Energies}

\author{Fagner C. Correia}
\email[Electronic address: ]{ ccorreia@ift.unesp.br}
\affiliation{J. Stefan Institute, Jamova 39, P. O. Box 3000, 1001
  Ljubljana, Slovenia}
\affiliation{Instituto de Física Teórica - Universidade Estadual Paulista, R. Dr. Bento Teobaldo Ferraz 271, Barra Funda Sao Paulo - SP, 01140-070, Brazil}

\author{Svjetlana Fajfer}
\email{ svjetlana.fajfer@ijs.si}
\affiliation{Department of Physics,
  University of Ljubljana, Jadranska 19, 1000 Ljubljana, Slovenia}
\affiliation{J. Stefan Institute, Jamova 39, P. O. Box 3000, 1001
  Ljubljana, Slovenia}

\begin{abstract}
We investigate  a spontaneously broken $U(1)_d$ gauge symmetry with a muon-specific dark Higgs. Our first goal is to verify how the presence of a new dark Higgs, $\phi$, and a dark gauge boson, $V$, can simultaneously face the anomalies from the muon magnetic moment and the proton charge radius. Secondly, by assuming that $V$ must decay to an electron-positron pair, we explore the corresponding parameter space determined with the low energy constraints coming from $ K \to \mu X$, electron $(g-2)_e$, $K \to \mu \nu_\mu  e^+ e^-$, $K \to \mu \nu_\mu  \mu^+ \mu^-$, $\tau \to \nu_\tau \mu \nu_\mu e^+ e^-$. We focus in the scenario where the $V$ mass is below $ \sim 2 m_\mu$ and the $\phi$ mass runs from few MeV till $250$ MeV, with V-photon mixing of the order $ \sim\mathcal{O}(10^{-3})$. Among weak process at low energies, we check the influence of the new light vector on kaon decays as well as on the scattering $e^+ e^- \rightarrow \mu^+ \mu^-  e^+ e^-$ and discuss the impact of the dark Higgs on $e^+ e^- \rightarrow \mu^+ \mu^- \mu^+ \mu^- $. Finally, we consider contributions of the V-photon mixing in the decays $\pi^0  \to \gamma e^+ e^-$, $\eta  \to \gamma e^+ e^-$, $\rho \to \pi e^+ e^-$, $K^* \to K e^+ e^-$ and $\phi (1020) \to \eta e^+ e^-$.
 
 
\end{abstract} 
\maketitle

The dark matter abundance in the universe has stimulated numerous searches for the Standard Model (SM) extensions. Nevertheless, the true nature of the supposed new interaction is not understood yet and many models were suggested to explain it by assuming the existence of cosmologically stable particles, ranging in mass from below 1 GeV to above 1 TeV \cite{Davoudiasl:2014kua}. 

Moreover, there is a number of discrepancies between SM theoretical predictions and  experimental results at energies below the kaon mass which might be signatures of new physics. The long lasting muon anomalous magnetic moment $(g-2)_\mu$  puzzle, for instance, is still present at $3.6\sigma$ level and as claimed by many authors \cite{Batell:2009yf,Batell:2011qq,Fayet:2007ua,Carlson:2013mya,Davoudiasl:2012qa,Davoudiasl:2014kua,Lee:2014tba,Karshenboim:2014tka,Batell:2016ove},
it can be explained by new dark bosons with the masses below 200 MeV. One more recent example is the discrepancy between the proton charge radius measured via the Lamb shift in atomic and muonic hydrogen (see e.g. \cite{Barger:2010aj,Carlson:2015jba,Carlson:2012pc}). 

One of the proposals to explain the origin of such low energy puzzles is based on the spontaneously broken $U(1)_d$ gauge symmetry  \cite{Batell:2009yf,Batell:2011qq,Fayet:2007ua,Carlson:2013mya,Davoudiasl:2012qa,Davoudiasl:2014kua,Lee:2014tba,Batell:2016ove,Karshenboim:2014tka}, 
introduced in the context of astrophysical anomalies (see e.g. \cite{ArkaniHamed:2008qn,Boehm:2003bt}). Its basic mechanism allows the gauge coupling to be $\sim O(10^{-3})$ and implies a kinetic mixing amplitude between the new gauge boson $V$ and the photon field.

In decays with particles identified through the missing energy one might expect that some set of the invisible states is due to the existence of the coupled dark sector. Many experiments are devoted to search for weakly interacting particles \cite{Essig:2013lka,TheBABAR:2016rlg,Lees:2014xha} and one pioneering work in this direction was done by the authors of \cite{Cable:1974nx}, who succeeded to put limits on the decay $BR(K \to \mu \ \text{missing energy})$. By applying these bounds, the authors in \cite{Barger:2011mt} have found, for example, that the leptonic decay $K \to \mu \nu V$ is already very constraining on the $V$ parameters. 

In this paper we re-investigate a spontaneously broken $U(1) _d$ gauge model following the ideas presented in \cite{Batell:2011qq} and implementing additional constraints. First we assume that both the dark gauge boson, $V$, and the dark Higgs, $\phi$, cannot be directly detected and assume that they both are present in the explanation of  the proton size anomaly and kaon leptonic decays. We find  a tension between the upper bounds on the decay width of the kaon leptonic decay  and the proton size band for  a specific range of  relevant parameters. Secondly, we loose  this prior restriction and since we are mainly interested in the low mass region, we continue to treat the dark Higgs as  the muon-specific scalar, contributing to the   missing mass and work in the scenario where $V$ must decay  to 
$ e^+ e^-$. The model will be further constrained by the BaBar additional observables: the uncertainty in $\Gamma(K \to \mu \nu_\mu  e^+ e^-)$, for $m_{ee} > 145$ MeV, the upper bound for $\tau \to \nu_\tau \mu \nu_\mu e^+ e^-$ and by the electron anomalous magnetic moment $(g-2)_e$. When the dark Higgs mass is in the range 
$ 2 m_\mu \leq m_\phi \leq (m_K -m_\mu)$, we  derive bounds from the experimental upper bound on $\Gamma(K \to \mu \nu_\mu  \mu^+ \mu^-)$. 
  
The analysis described above can be considered complementary to the recent BaBar result \cite{TheBABAR:2016rlg} on the search for a new neutral vector boson in the process $e^+ e^- \to \mu^+ \mu^- V$. Their result has placed very strong limits on the coupling constant of $V$, indicating that the presence of a massive vector state can be excluded in the range $0.212- 10$ GeV.

Section I contains the description of the model we explore in our study. Sec. II is devoted to the derivation of bounds from low energy phenomenology. In Sec. III we discuss implications of this proposal on the low energy processes and Sec. IV contains the short summary of our results. 
  
\section{Framework: Dark ${\rm U(1)_d}$ }

The $ U(1)_d$ gauge invariant Lagrangian under consideration is written by \cite{Batell:2011qq}:
\begin{equation}
{\cal L} = -\frac{1}{4} V_{\alpha\beta} V^{\alpha\beta} +|D_\mu \phi|^2 +\bar \mu_R i \cancel{D} \mu_R -\frac{\kappa}{2} V_{\alpha \beta} F^{\alpha \beta} -  \bar L \mu_R H_{SM} \frac{\phi}{\Lambda} + h.c.
\label{e1}
\end{equation}
Here V is the gauge boson, neutral under the SM gauge group and charged under $U(1)_d$. The field  $\phi$ is the dark Higgs with a condensate $\langle\phi\rangle = \frac{v_R}{\sqrt{2}}$. The covariant derivative $D_\alpha = \partial_\alpha + i g_R V_\alpha + i e Q_{EM} A_\alpha$ and $\kappa$ is the mixing angle. 

The muon mass is then introduced as $v v_R/ ({2} \Lambda)$, while the SM-like Yukawa coupling is given by $v_R/ ({\sqrt 2} \Lambda)$. As asserted by the authors of \cite{Batell:2011qq}, the proton  charge radius phenomenology will favor the range of the new parameters such that the scale $\Lambda$ can be at the weak scale. Moreover, the model given in (\ref{e1}) leads to gauge anomalies involving the photon and the vector $V$ and in order to restore gauge invariance, it is mandatory to introduce new dynamical scalar degrees of freedom. 

There are different ways to make this theory UV complete. For instance, a number of SM extensions with new vector-like fermions were constructed for this purpose \cite{Goertz:2015nkp,FileviezPerez:2010gw,Chao:2010mp,Ko:2010at,Duerr:2013dza,Schwaller:2013hqa}. In \cite{Chen:2015vqy} it was suggested to extend the $\text{SM}\otimes U(1)_d$  by three right-handed neutrinos in order to generate neutrino masses. One last example was recently offered in \cite{Batell:2016ove} by  the  "lepton-specific" representation of a generic two Higgs doublet model in which the scalar sector contains the SM Higgs, an additional doublet and the dark $\phi$. 

The Lagrangian given in (\ref{e1}) leads to the following couplings of the new vector and scalar to fermions:
\begin{eqnarray}
V \rightarrow -i \gamma^\mu(\bar{g}^\mu_V + \bar{g}^\mu_A\gamma_5), \qquad
\phi \rightarrow -i g_\phi
\end{eqnarray}
with the definitions
\begin{eqnarray}\label{gsgr} 
\bar{g}^\mu_V &=& e \kappa + \frac{g_R}{2}, \quad
\bar{g}^\mu_A = \frac{g_R}{2}, \quad
g_\phi = g_R \frac{m_\mu}{M_V} 
\end{eqnarray}

In our analysis we choose to work with the set of parameters $(g_\phi, m_\phi, \kappa, M_V)$, by assuming $g_R = 2 \lambda \kappa$ which, from the relation (\ref{gsgr}), leave us with: 
\begin{equation}\label{lambdadef}
\lambda = \frac{M_V}{m_\mu} \frac{g_\phi}{2 \kappa}.
\end{equation}
In the next section we will preferably consider specific choice of $(\lambda, m_\phi)$ since in the literature bounds and predictions are often presented for the space $(M_V, \kappa)$. We will also mention this combination in Sec. III.

\section{Low energy phenomenology bounds}

One of  main goals  of the model  presented in eq.~(\ref{e1}) was  to explain the proton size discrepancy \cite{Batell:2011qq}. It was first noticed by the authors of \cite{Barger:2011mt} that the $K \to \mu X$ decay, with $X$ being a set of states seen only as missing energy, can give very strong constraints on the parameters of $V$ \cite{Cable:1974nx}. Nevertheless, it was assumed that only an invisible vector state gives new contribution to this process. We first try  to establish  parameter space of  $(M_V, \kappa)$  which is allowed by the proton charge radius and the leptonic kaon decay, including the contributions of both vector and scalar dark bosons as missing mass.

The procedure described above can be summarised as:
\begin{itemize}
\item \textit{Proton Charge Radius}

The measurement of the Lamb shift in muonic and atomic hydrogen (\cite{Antognini:2013jkc,Pohl:2010zza}) has indicated a difference for the proton radius square, $r_p^2$, which can be abbreviated to (for details see \cite{Batell:2011qq}):
\begin{equation}\label{PR1}
\Delta r^2_p = (r_p)^2_{e-p} - (r_p)^2_{\mu-p}=0.060(12) \ \text{fm}^2.
\end{equation}
As discussed in \cite{Batell:2011qq}, this discrepancy can be properly approached by the model of eq. (\ref{e1}) due to the mixing with the photon. Moreover, since the dark Higgs couples to  muons only, the mass $m_\phi$ will remain free to adjust additional limits. Here we rewrite the theoretical corrections to the difference  in eq.(\ref{PR1}), following the notation of eq. (\ref{gsgr}):
\begin{equation}
\Delta r^2|_{e-H} = - \frac{6 \kappa^2}{M_V^2}, \qquad \Delta r^2|_{\mu-H} = - \frac{6 \kappa^2 (1 + \frac{\lambda}{e})}{M_V^2}   f(a M_V),
\label{pcr}
\end{equation}
where $ a = (\alpha m_\mu m_p)^{-1} (m_p  +m_\mu)$ is the $\mu-H $ Bohr radius, $\alpha$ is the fine-structure constant, and $f(z) = (z/(1+z))^4 $. 
Therefore, for  $a M_V \gg 1$ one can obtain the 2$\sigma$  favourable region  for the parameter $\kappa$, using the proton radius discrepancy given in eq. (\ref{pcr}):

\begin{eqnarray}\label{pcr-reg}
\kappa^2 =  \frac{e M_V^2}{6 \lambda}( \Delta r_p^2 \pm 2 \sigma).
\end{eqnarray}

\item \textit{Muonic Kaon Decay $K \to \mu X$} 

In the context of $V$ and $\phi$ bremsstrahlung from $\mu$, the result of analysis in ref. \cite{Cable:1974nx} can be converted to the upper bound:
\begin{eqnarray}
\frac{\Gamma_{K\rightarrow \mu X}}{\Gamma_{K\rightarrow \mu \nu}} < 3.5 \times 10^{-6},
\qquad 227.6 < m_X (\text{MeV}) < 302.2 \qquad   90\% \ C.L.
\label{Pangcond}
\end{eqnarray}
with
\begin{eqnarray}
\Gamma_{K\rightarrow \mu X} &=& \Gamma_{K\rightarrow \mu \nu V} + \Gamma_{K\rightarrow \mu \nu \phi} 
\label{calcPang1}
\end{eqnarray}
Note that in eq. (\ref{Pangcond})  there is an experimental acceptance on the missing mass, $m_X$. The eq. (\ref{calcPang1}) can be written as function of $(\kappa, M_V, \lambda, m_\phi)$.
\end{itemize}

\begin{figure}[h]
	\centering
	\subfloat[]{\includegraphics[width=.45\columnwidth]{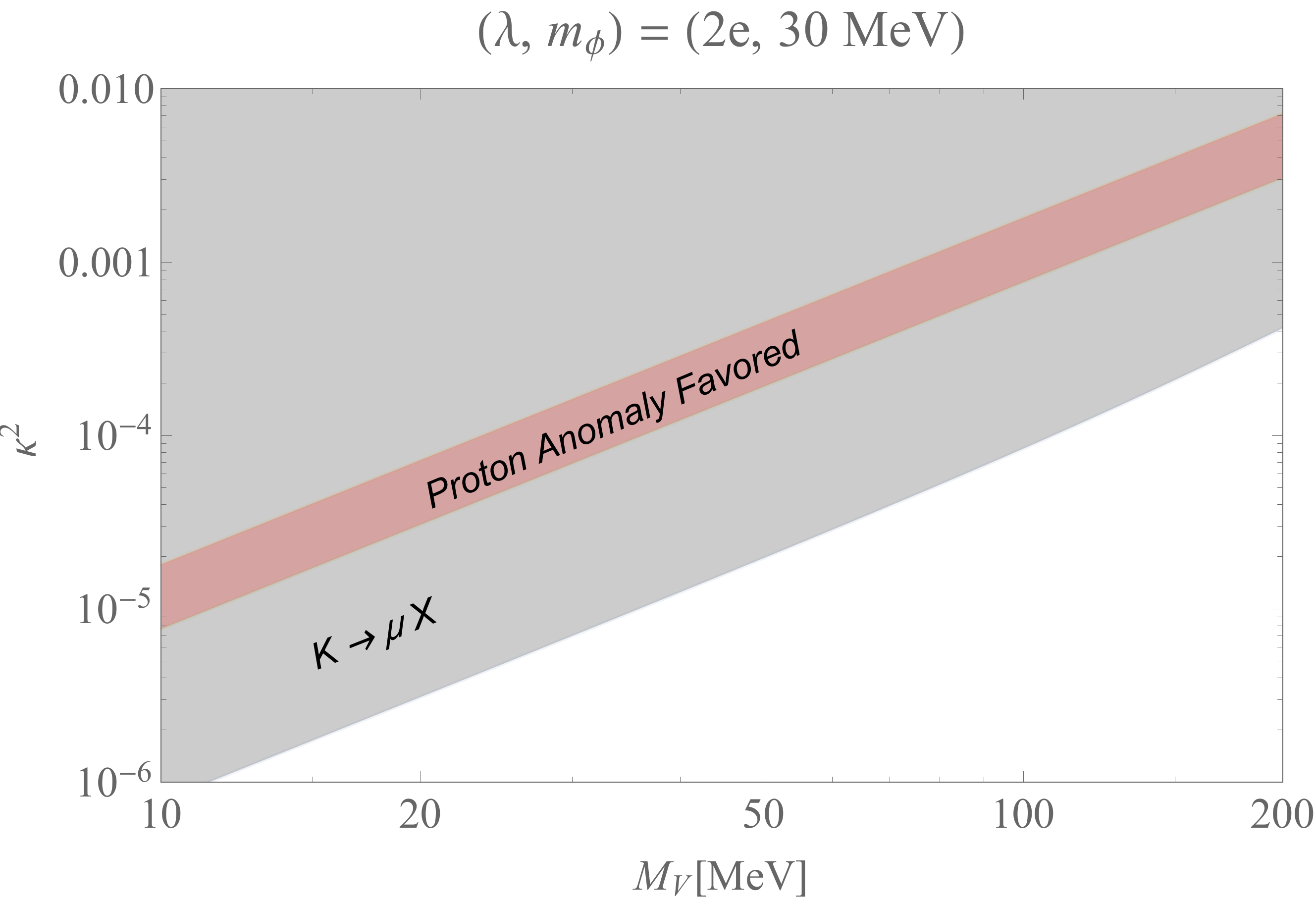}} \qquad
	\subfloat[]{\includegraphics[width=.45\columnwidth]{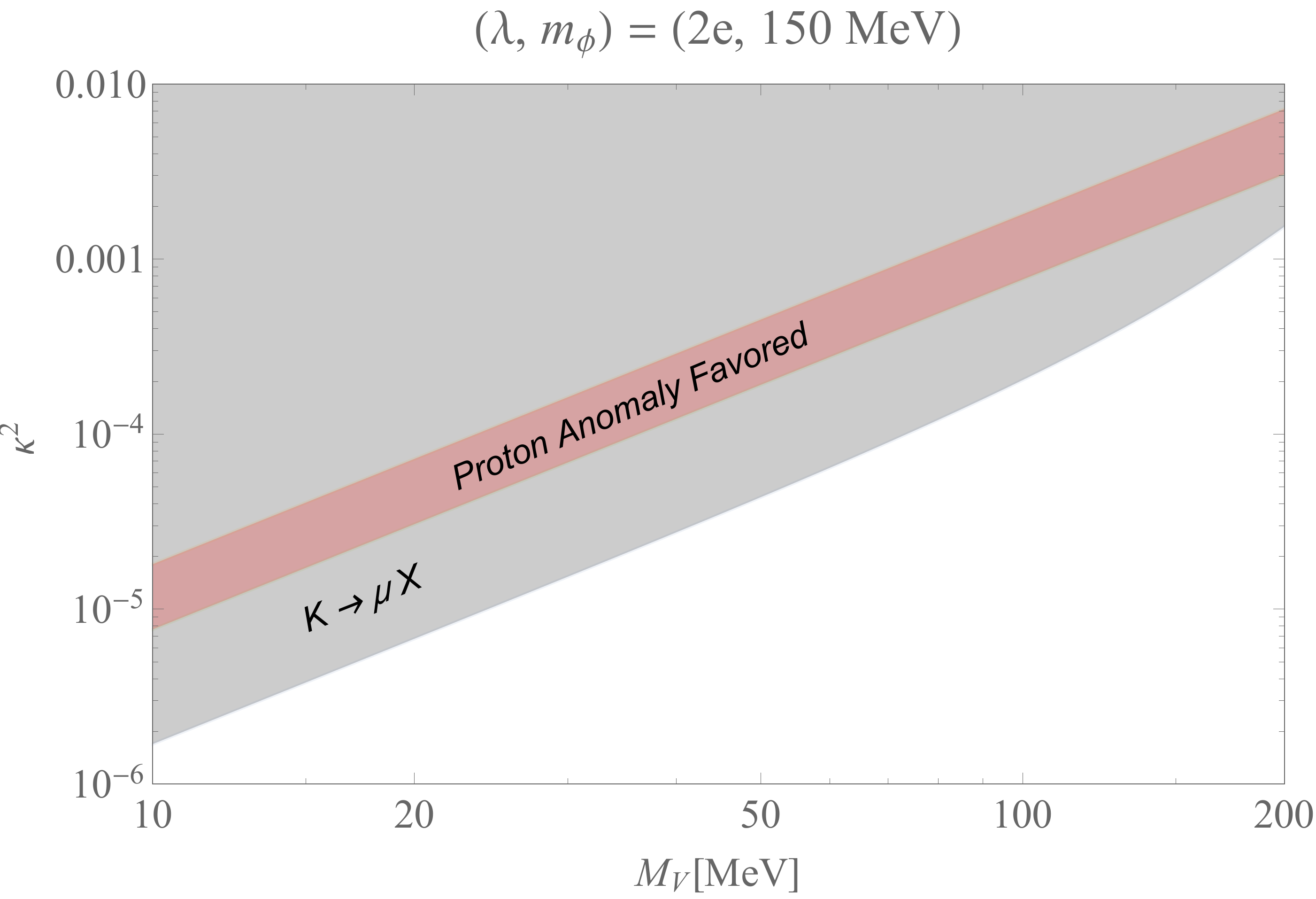}} 
	\caption[]{The $(M_V, \kappa^2)$ parameter space fixed by the bounds from $K \to \mu X$ with the muonic  $V$ and $\phi$ bremsstrahlung (grey) along with the allowed area of the proton size anomaly (pink). The grey color marks the excluded region at the $90 \%$ C.L.} 
	\label{fig:first}
\end{figure}

In Fig.  (\ref{fig:first})  we present  the allowed parameter space $(M_V,\kappa^2)$ obtained when the constraints from eqs. (\ref{pcr-reg}) and (\ref{Pangcond}) are applied, for fixed values of $(\lambda, m_\phi)$. On these particular examples the grey colour denotes the region  excluded at the $90 \%$ C.L. by the bound on $BR(\Gamma(K\rightarrow \mu X)$, while the pink one denotes the region allowed by the proton size anomaly up to $2 \sigma$. 
We have checked if there is  any region   which can satisfy both conditions.  For a large set of the $(\lambda, m_\phi)$  points we could not find any positive solution.

In this section we would like to illustrate how the muonic kaon decay  itself is very restrictive. Once the dark Higgs is muon-specific, and we are mainly interested in the regime of small masses (bellow $2 m_\mu$), we have to relax  our first assumption and assume that the gauge boson $V$  decays to electron-positron pair $V \to e^+ e^-$. 
However, such $V e^+ e^-$ interaction  creates additional effects in a number of processes.

\subsection{Muon Anomalous Magnetic Moment}

The discrepancy between experimental results and the SM prediction for $(g-2)_\mu$ persists as an intriguing low energy puzzle in particle physics, currently being $ \delta a_\mu^{\text{exp-SM}}  =\Delta  (g-2)_\mu/2 = 288(80) \times 10^{-11}$ \cite{Agashe:2014kda}.

The model of eq. (\ref{e1}) contributes at one-loop level with the three different contributions to $\rm{(g-2)_\mu}$ - vector, axial-vector and the scalar one. 
The authors of  \cite{Batell:2011qq} have noticed that within this framework there will be an enhancement of opposite sign to $\delta a_\mu$, if compared to the pure vector case. Such feature  might allow an overlap with the  proton anomaly allowed region, since in the eq. (\ref{pcr-reg}) there is no dependence on $m_\phi$. We can clearly see this feature through writing the complete expression below (see \cite{Leveille:1978px} and  \cite{McKeen:2009ny}):
\begin{eqnarray}\label{gmufunction}
		\delta a_\mu &=& (g_V)^2 I_V\Bigl(\frac{M_V^2}{m_\mu^2}\Bigr) + (g_A)^2 I_A\Bigl(\frac{M_V^2}{m_\mu^2}\Bigr) + (g_\phi)^2 I_\phi\Bigl(\frac{m_\phi^2}{m_\mu^2}\Bigr), \nonumber\\
	          &=& \kappa^2 \Bigl[(e + \lambda)^2 I_V\Bigl(\frac{M_V^2}{m_\mu^2}\Bigr) + (\lambda)^2 I_A\Bigl(\frac{M_V^2}{m_\mu^2}\Bigr) + \Bigl(\frac{2 m_\mu \lambda}{M_V}\Bigr)^2 I_\phi\Bigl(\frac{m_\phi^2}{m_\mu^2}\Bigr) \Bigr]  \nonumber\\
	          &\equiv& \kappa^2 F(\lambda,M_V,m_\phi).
\end{eqnarray}

The  full expression for the one-loop integrals $I_{V,A,\phi}$ can be found in  \cite{Leveille:1978px}. We note that in the regime where the function $F(\lambda, M_V, m_\phi)$ is small the coupling $\kappa$ can reach arbitrarily large values. This behaviour will be tested in the subsection \ref{discuss}.

\subsection{Leptonic Kaon Decays}

The first conclusion  of our analysis is that the dark photon must decay to an electron-positron pair. This, however,  does not mean that $K \to \mu X$ bound is not important anymore, since the scalar $\phi$ still takes a role as a missing mass.  The relation (\ref{gsgr}) accompanied by $g_R = 2 \lambda \kappa$, will again produce the  excluded region for $(M_V, \kappa^2)$,  even being dependent on $m_\phi$. Moreover, given the richness of  kaon phenomenology, the new requirement applied on the additional channels involving $V$ can produce even stronger bounds on the parameter space:

\begin{itemize}
\item $K^+ \rightarrow \mu^+ \nu_\mu e^+ e^-$ 

The branching ratio for this process is given in  \cite{Agashe:2014kda}:
\begin{equation}\label{kmuee}
\frac{\Gamma(K^+ \to \mu^+ \nu_\mu e^+ e^-)}{\Gamma_K} = 7.06(31) \times 10^{-8} \quad (m_{ee} > 145 MeV).
\end{equation}

The authors of \cite{Carlson:2013mya} considered contributions of $V$ via kinetic mixing with a radiated SM photon, as in refs. \cite{Carlson:2012pc,Bijnens:1992en,Cirigliano:2007xi}. They also made a comparison  between $K^+ \rightarrow \mu^+ \nu_\mu V \rightarrow \mu^+ \nu_\mu e^+ e^-$ and the QED background \cite{Poblaguev:2002ug}, having found that a new light vector boson, if it  decays before leaving the detector, might produce bumps in the electron-positron invariant mass spectrum. The model  presented in eq. (\ref{e1}), which we  use, leads to  a signal few orders of magnitude larger than the proposal of \cite{Carlson:2013mya}.

We assume that these NP corrections by itself should not be larger than $1\sigma$ of the result given in eq. (\ref{kmuee}). Using  the narrow-width approximation and anticipating that  the range  for $M_V$ is $145$ MeV $ < M_V  < 2m_\mu$, where $V$ can decay only to $e^+ e^-$, we can impose the following upper bound: 
	\begin{eqnarray}
	\Gamma(K^+ \rightarrow \mu^+ \nu_\mu V, V \rightarrow e^+ e^- ) = \Gamma(K^+ \rightarrow \mu^+ \nu_\mu V) \times \text{Br}(V \rightarrow e^+ e^-).
	\end{eqnarray}
Finally, since $\text{BR}(V \rightarrow e^+ e^-) = 1$ it becomes:
	\begin{eqnarray}
\frac{\Gamma(K^+ \rightarrow \mu^+ \nu_\mu V)}{\Gamma_K} < 3.1 \times 10^{-9}.
	\end{eqnarray}
 
\item $K^+ \rightarrow \mu^+ \nu_\mu \mu^+ \mu^-$

 If the muon-specific dark Higgs has a mass larger than $2m_\mu$, the $\phi$ bremsstrahlung will be then followed by  the decay $\phi \to \mu^+ \mu^-$. 
 In this case the bound from the  $K \to \mu X$ cannot be applied.  The $K \to \mu X$  constraint should be replaced by the existing  upper bound \cite{Agashe:2014kda}: 
	\begin{eqnarray}\label{k3mu}
\frac{\Gamma(K^+ \rightarrow \mu^+ \nu_\mu \mu^+ \mu^-)}{\Gamma_K} < 4.1 \times 10^{-7}, \qquad 90\% \ C.L.
	\end{eqnarray}
The above expression can be useful in the region $2m_\mu < m_\phi < (m_K- m_\mu)$ and  as we will find out  in the section \ref{discuss}, if $m_\phi$ is close to $2m_\mu$  this constraint is equally powerful as one coming from  $K \to \mu X$ (\ref{Pangcond}). Since  the dark Higgs interacts with muons only, its decay to $\mu^+ \mu^-$ is allowed now and $\text{Br}(\phi \rightarrow \mu^+ \mu^-) = 1$.  Using  the narrow-width approximation,  we obtain:
\begin{eqnarray}
\frac{\Gamma(K^+ \rightarrow \mu^+ \nu_\mu \phi)}{\Gamma_K} \times \text{BR}(\phi \rightarrow \mu^+ \mu^-) = \frac{\Gamma(K^+ \rightarrow \mu^+ \nu_\mu \phi)}{\Gamma_K} < 4.1 \times 10^{-7}, \qquad 90\% \ C.L.	
\end{eqnarray}
\end{itemize}

\subsection{Constraints from  $\tau^+ \to \nu_\tau \mu^+ \nu_\mu e^+ e^-$}

Within the SM, the $e^+ e^-$ pair in the process $\tau^+ \to \nu_\tau \mu^+ \nu_\mu e^+ e^-$ originates from the virtual photon or Z emission in the decay 
$\tau^+ \to \nu_\tau \mu^+ \nu_\mu $. Analogously to the previous  case  with $ V\to e^+ e^-$, we can consider the upper bound to the ratio of this process by assuming that one can safely use the narrow-width approximation: 
\begin{eqnarray}\label{tau}
\frac{\Gamma(\tau \rightarrow \nu_\tau \mu \bar{\nu}_\mu V )}{\Gamma_\tau} \text{BR}(V \rightarrow e^+ e^-) = \frac{\Gamma(\tau \rightarrow \nu_\tau \mu \bar{\nu}_\mu V )}{\Gamma_\tau} < 3.6 \times 10^{-5}, \qquad 90\% \ C.L.
\end{eqnarray}		

The differential decay rate for $\tau^+ \to \nu_\tau \mu^+ \nu_\mu e^+ e^-$ is given by:
		\begin{eqnarray}
		d\Gamma_{\tau \rightarrow \nu_\tau \bar{\nu}_\mu V \mu} &=& \frac{m_\tau^3}{256 (2 \pi)^6} |\mathcal{M}_{\tau \rightarrow \nu_\tau \bar{\nu}_\mu V \mu}|^2 \sqrt{\lambda(1,\delta_3,0)} \sqrt{\lambda(\delta_2,\delta_\mu,\delta_V)} \nonumber \\ & & \times \frac{\sqrt{\lambda(\delta_3,\delta_2,0)}}{\delta_2 \delta_3} d\delta_2 d\delta_3 dC_{\theta_2} dC_{\theta_3} d\phi,
		\end{eqnarray}
		where we have assigned the momenta $\tau(k), \mu(p_1), V(p_2), \nu_\mu(p_3), \nu_\tau(p_4)$. $C_{\theta_2}$ is the angle between $\mu$ and $\tau$ momenta in the rest frame of $k_2 \equiv p_1 + p_2$ and $C_{\theta_3}$ is the angle between $k_2$ and $k$ in the rest frame of $k_3 \equiv p_1 + p_2 + p_3$. Besides, $\delta_i \equiv \frac{M_i^2}{m_\tau^2}$, $i = \mu, V, 2, 3$ and $M_j^2 \equiv k_j^2$,  $j = 2, 3$, $\phi$ is the angle between the planes composed by $\vec k_1 \times \vec k_2$ and $\vec k_2 \times \vec k_3$ and $\lambda(a,b,c) = \bigl(a - (\sqrt{b} + \sqrt{c})^2\bigr)\bigl(a - (\sqrt{b} - \sqrt{c})^2\bigr)$.  

We point out  that the same analysis could be done for $\mu \rightarrow \nu_\mu e \bar{\nu}_e e^+ e^- $, but this is not as restrictive as the constraint (\ref{tau}),  due to the  smaller phase-space. 
	
\subsection{Electron Anomalous Magnetic Moment}

The Standard Model contribution to $a_e = \frac{(g-2)_e}{2}$ has been recently improved up to the tenth order, corresponding to $\Delta a_e = 1 159 652 181.78 (77) \times 10^{-12}$ \cite{Aoyama:2012wj}, facing the experimental value $a_e= (g-2)/2 = (1159.65218076\pm 0.00000027)\times 10^{-6}$ \cite{Agashe:2014kda}.
In \cite{Pospelov2009} the author argues that the one-loop correction to this quantity must be reinterpreted as an effective shift of the fine-structure constant, which would not exceed 15 ppb (see eq.(6) of \cite{Pospelov2009}), leading to  the following constraint:
		\begin{eqnarray}
		(e \kappa)^2 I_V\Bigl(\frac{M_V^2}{m_e^2}\Bigr) < 1.5 \times 10^{-8}.
		\end{eqnarray}
The above relation will be considered along with all the bounds presented in the previous subsections.

\subsection{Experimental bounds}

There are many experimental searches for the dark sector (see e.g \cite{Essig:2013lka,Babusci:2014sta,Merkel:2014avp,delRio:2016anz,Lees:2014xha,Batley:2015lha,Anastasi:2015qla,::2016lwm,Archilli:2011zc,Bjorken:1988as}).
We mention here only the most recent bounds. The NA48/2 collaboration \cite{Batley:2015lha} has searched for bound in $\pi^0 \to \gamma e^+ e^-$ decay and obtained that 
$\kappa^2 =(0.8-1.11)\times10^{-5}$ at $90\%$ C.L.  for the mass of the vector gauge boson in the range $2 m_e < M_V < 140$ MeV. 
The Kloe-2 collaboration determined  the bound on the mass of dark photon and photon-dark photon mixing parameter from the study of dark photon contribution  in the $\phi \to \eta V \to \eta e^+ e^-$ decay, by measuring 
the cross sections $e^+ e^- \to V \gamma \to  \mu^+ \mu^-  \gamma $ and $e^+ e^- \to V \gamma \to e^+ e^-  \gamma$. They found  that $\kappa^2$ has to be smaller then $5 \times 10^{-5}$ \cite{DeSantis:2015cma,delRio:2016anz}. 

The  BaBar collaboration   obtained the very restrictive bounds  on the dark $Z^\prime$ boson  (corresponds to $V$ in our case) \cite{TheBABAR:2016rlg} from the cross section for the  $e^+ e^- \to \mu^-\mu^+ Z^\prime\to \mu^+ \mu^-\mu^+ \mu^-$ process relying on the model described in  \cite{He:1990pn,He:1991qd}. A basic feature of this model is the absence of the $Z^\prime$  coupling to the first lepton generation.
The BaBar search is based on 514 fb$^{-1}$  of data collected at the PEP-II 
$e^+ e^-$ storage ring, predominantly taken at the $\Upsilon (4s)$ resonance and their result is applicable also on the models in which the gauge bosons are coupled exclusively to right-handed muons. They obtained a strong bound on the coupling and the mass of $Z^\prime$  in the region $0.2$ GeV $< m_{Z'} < 4$ GeV.  Since our model contains both dark bosons,  dark Higgs and dark gauge boson, we also expect that the inclusion of the dark Higgs contribution in  $e^+ e^- \to \mu^+ \mu^-\mu^+ \mu^-$ might  only slightly modify the phenomenology of this channel. Thus, we combine the results of NA48/2, Kloe-2 and BaBar  analysis on our plots,  which we present in the next section.

\begin{figure}
	\centering
	\subfloat[For $(\lambda,m_\phi) = (0.8e, 30 \text{MeV})$ the search for missing mass in $K \to \mu X$  excludes the $\pm 2\sigma$ favored region for the proton anomaly.]{\includegraphics[width=.45\columnwidth]{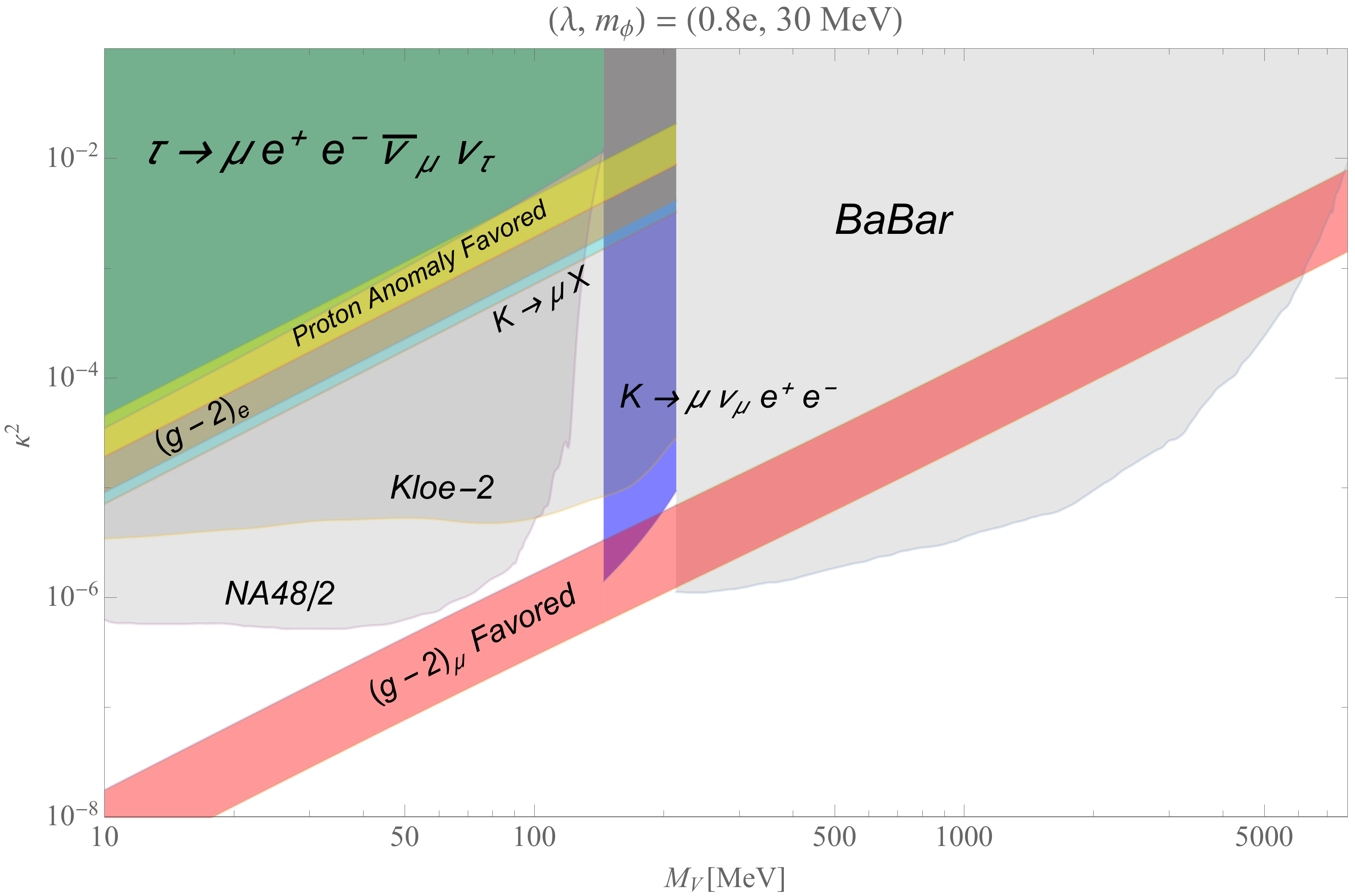}} \qquad
	\subfloat[For $(\lambda,m_\phi) = (0.8e, 150 \text{MeV})$ the constraint from $BR(K \to \mu X)$  can be relaxed, allowing a small  $\pm 2\sigma$ region to accommodate the proton anomaly, which is however excluded by $(g-2)_e$ region. There is no overlap between regions allowed by the  proton and $(g-2)_\mu$ discrepancies.]{\includegraphics[width=0.45\columnwidth]{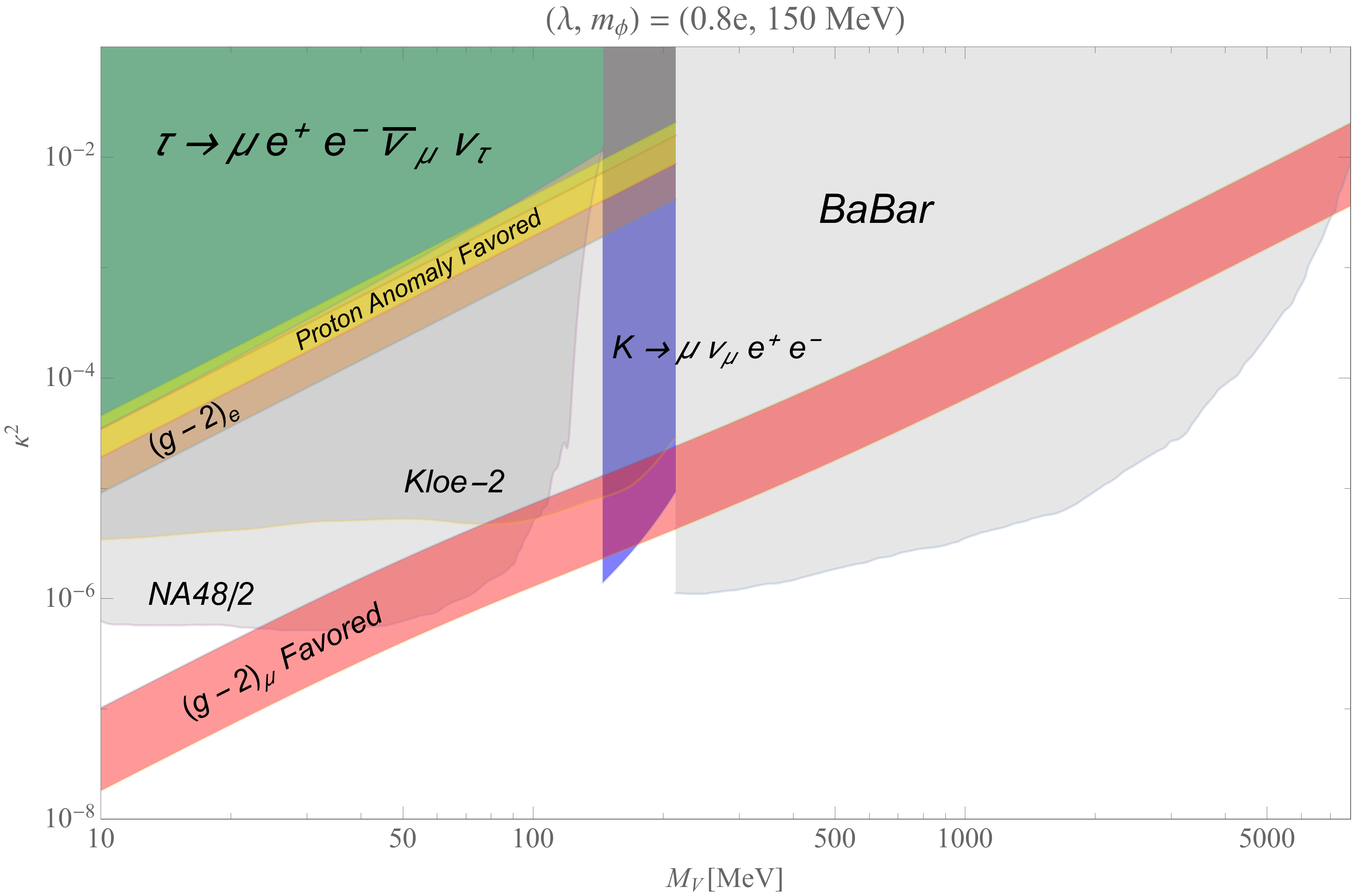}} \qquad
	\subfloat[For $(\lambda,m_\phi) = (2e, 30 \text{MeV})$ the bound  for $BR(K \to \mu X)$  will exclude at  $\pm 2\sigma$ level, the favored region for proton anomaly explanation. ]{\includegraphics[width=.45\columnwidth]{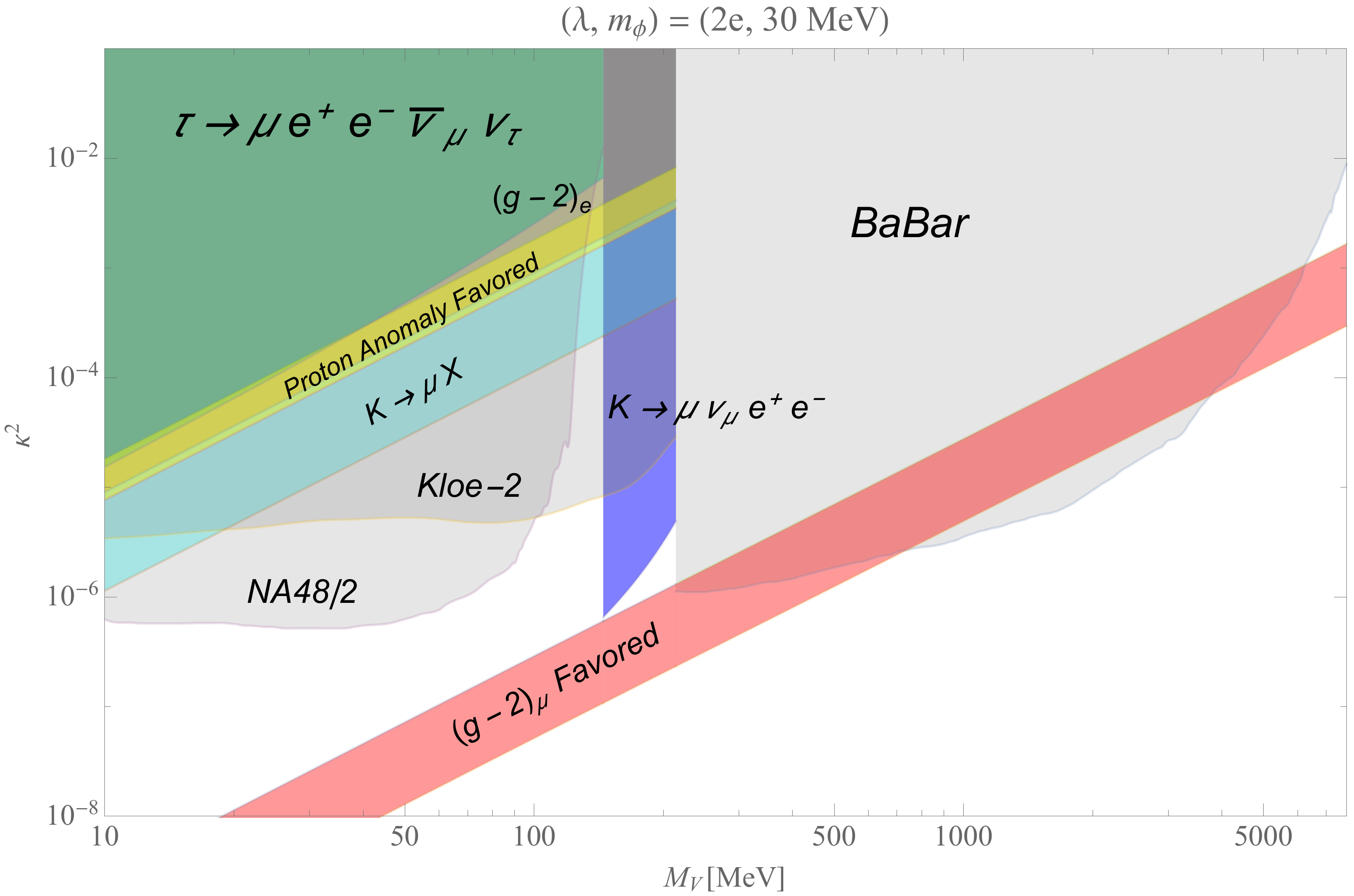}} \qquad
	\subfloat[For $(\lambda,m_\phi) = (2e, 150 \text{MeV})$ the leptonic kaon channel  and the 2016 BaBar data start to exclude the $\pm 2\sigma$ favored region for $(g-2)_\mu$.]{\includegraphics[width=.45\columnwidth]{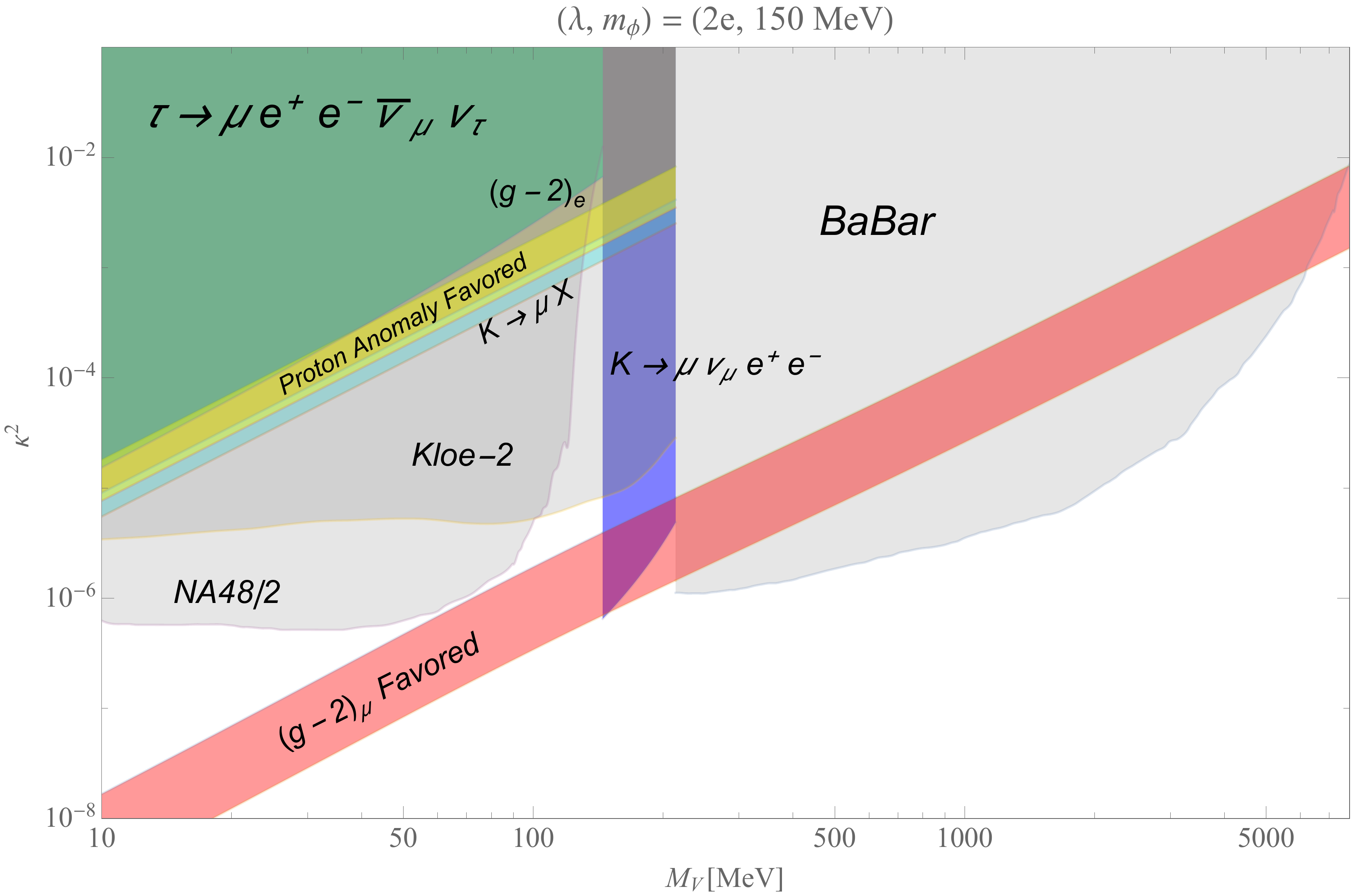}} 
	\caption[New Constraints.]{Parameter space for $(M_V, \kappa^2)$. The colored regions are excluded by the respective processes and the region  favored  by $(g-2)_\mu$ at $\pm 2 \sigma$  is marked by pink, while the region allowed by the  proton size anomaly  is  yellow.} 
	\label{fig:PS}
\end{figure}

\begin{figure}[h]
	\centering
	\includegraphics[width=10cm,clip=true]{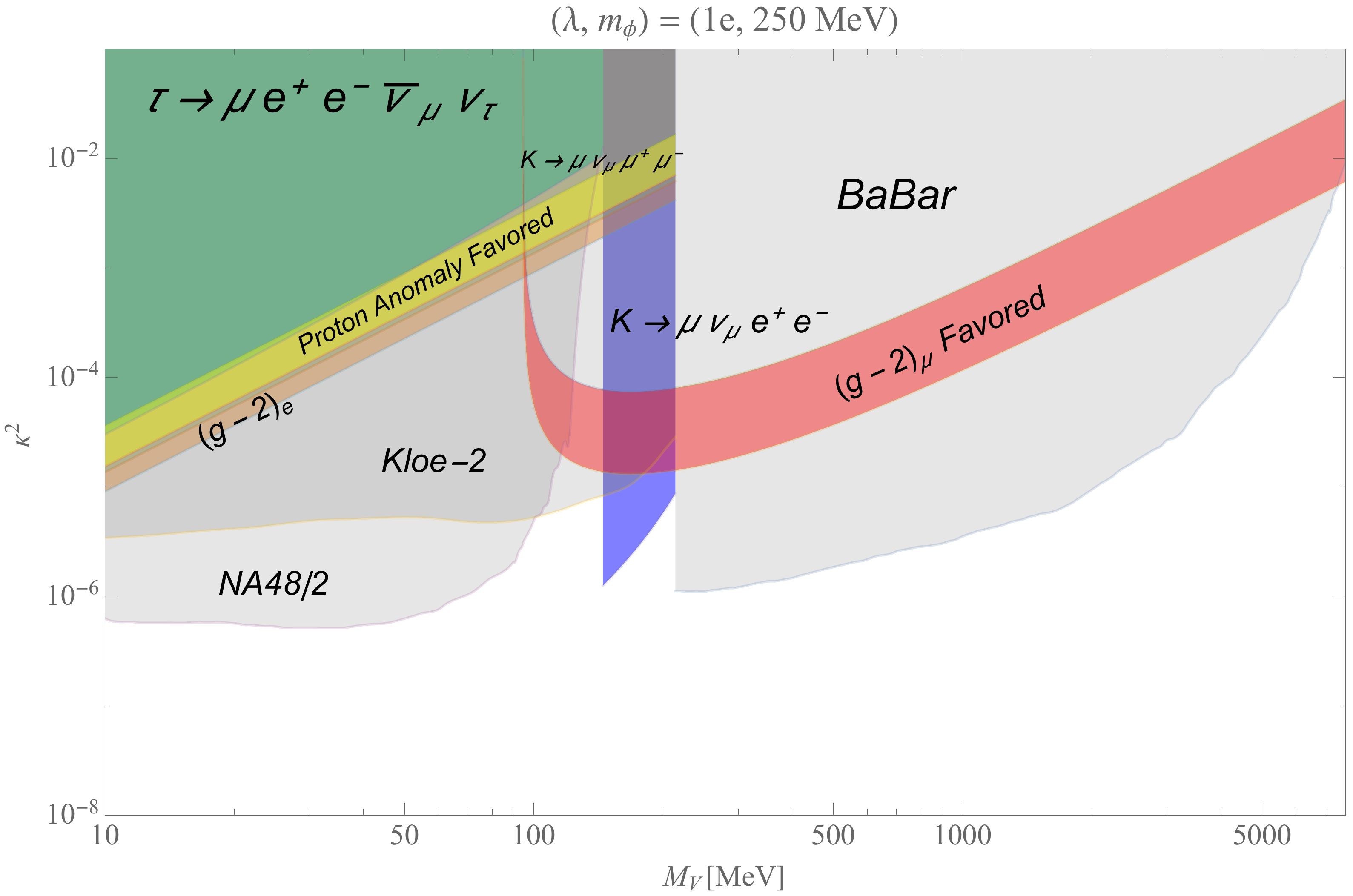}
	\caption{Parameter space for $(M_V, \kappa^2)$. The colored regions are excluded by the processes denoted on the respective areas. The region favored by $(g-2)_\mu$  at $\pm 2 \sigma$ is marked by  red and proton radius anomaly by yellow color.  The  size  of $\lambda$ and the large scalar mass are such that $F(\lambda,M_V,m_\phi)$ in eq. (\ref{gmufunction}) goes to zero, requiring large values of $\kappa$. For $m_\phi > 2m_\mu$ we use the constraint from $BR(K\rightarrow \mu \nu \mu \mu) < 4.1 \times 10^{-7}$ which excludes the proton favored region at  $90 \%$ C.L.}\label{fig:overlap2}
\end{figure}

\subsection{Discussion}\label{discuss} 
The constraints derived in the previous subsections are presented in Fig. \ref{fig:PS}. The colored areas are excluded, while the yellow and red bands correspond to the allowed region of the proton charge radius and muon anomalous magnetic moment at $2\sigma$ level, respectively. The regions excluded by NA48/2 \cite{Batley:2015lha}, Kloe-2 \cite{DeSantis:2015cma,delRio:2016anz} and BaBar  \cite{TheBABAR:2016rlg} are  grey.

In order to derive a more  general conclusion on the proton anomaly explanation, we can combine the definition in (\ref{lambdadef}) along with the constraint in (\ref{pcr-reg}) to obtain: 
\begin{equation}\label{lambda}
\lambda = \frac{3}{2 e} \frac{g_\phi^2}{\Delta r_p^2 \, m_\mu^2}. 
\end{equation}

In addition, by writing the amplitude of $K \to \mu \nu \phi$ as function of $(g_\phi, m_\phi)$ we have concluded that the parameter space for $g_\phi > 0.03$, with $m_\phi < 2 m_\mu$, will be ruled out at $90 \%$ C.L. The eq.(\ref{lambda}) translates this assertion to  $\lambda > 0.86 e$, a condition that can necessarily exclude the central value of the proton radius discrepancy. We can verify this, for instance, in Fig.\ref{fig:PS}. In (a), even if the above-mentioned limit is  respected, the dependence on the small $m_\phi$ results in the exclusion of the yellow region from the kaon muonic decay.  In Fig.\ref{fig:PS} (b) the $\lambda = 0.8 e$ accompanied with a large $m_\phi = 150$ $\, \text{MeV}$ will loose the $K \to \mu X$ bound. However, in both cases the bound from $(g-2)_e$ will disfavour this sector. In (c) and (d), since $\lambda = 2 e$, the proton band is necessarily excluded.

We stress that the only areas in Fig. \ref{fig:PS} dependent on $m_\phi$ are those related to $K \to \mu X$ and $(g-2)_\mu$. The four plots are pointing out that the proton charge radius cannot be explained by the spontaneously broken  dark $U(1)_d$  gauge symmetry. The whole region which allows to explain proton radius puzzle is being excluded by the constraints from $(g-2)_e$, $K \to \mu X$,
$K \to \mu \nu_\mu e^+ e^-$, and $\tau \to \mu \nu_\mu e^+ e^-$. 
In any of the cases we analyse, the proton charge radius anomaly and the muon anomalous magnetic moment cannot  be simultaneously explained. Apart from that, the constraint coming from  $K \to \mu \nu_\mu e^+ e^-$ can almost extrapolate the BaBar bound on the mass of vector gauge boson down to $M_V \sim 145$ $\, \text{MeV}$.  

If the size of $\lambda$ and a large scalar mass are such that the function $F(\lambda,M_V,m_\phi)$ in eq. (\ref{gmufunction}) becomes  very small, approaching to zero, the coupling $\kappa$ tied to the muon anomaly  might be  arbitrarily large. In Fig.\ref{fig:overlap2}, for example, the dark Higgs mass is $m_\phi = 250$ MeV and we can find a tiny overlap between the pink  and yellow bands. Nevertheless, for $m_\phi > 2m_\mu$ using the constraint $BR(K\rightarrow \mu \nu \mu \mu) < 4.1 \times 10^{-7}$, we find out that  it again leads to the exclusion of the proton favoured region at $90 \%$ C.L.

We finally note that the contribution of both the dark Higgs and the dark $V$ can enable $(g-2)_\mu$ to be explained. If, for example, only the dark gauge boson is present, there would be no region on the parameter space allowed by experimental results - and by the bounds we have shown here - that could explain the respective anomaly.

\section{Predictions at low energies}

Our analyses of the dark $U(1)_d$ gauge sector allows the mass of $M_V$ to be in the region around $50 < M_V (\text{MeV}) < 150$ with the parameter $\kappa \sim 10^{-3}$. 
One would expect that the weak decays are more likely to offer good testing ground for the dark sector \cite{Kamenik:2011vy}. Particularly,  the flavor changing neutral current processes occurring in meson decays were most favorable for such searches. For example in ref.  \cite{Davoudiasl:2012ig} the rare decay of K and B mesons to $\pi e^+ e^-$ were suggested  as  interesting candidates for  the dark boson searches, mainly due to the low rate of $BR(K^+ \to \pi^+ e^+ e^-)_{exp} = (3.00 \pm 0.09) \times 10^{-7}$ in the SM. By relying on the reanalysis of $K \to \pi  \gamma^*$ in \cite{Davoudiasl:2014kua}, we calculate the branching ratio for $K\to \pi V$ and present our result in Fig. \ref{fig:kpiV}. 
The $V$ dark  boson  promptly decays to $V \rightarrow e^+ e^-$ and the narrow-width approximation   will then give that 
$BR(K \rightarrow \pi  V \rightarrow \pi   e^+ e^-) = BR(K \rightarrow \pi  V )$.  

\subsection{$K \rightarrow \pi V$}

\begin{figure}[h]
	\centering
	\subfloat{\includegraphics[width=0.65\columnwidth]{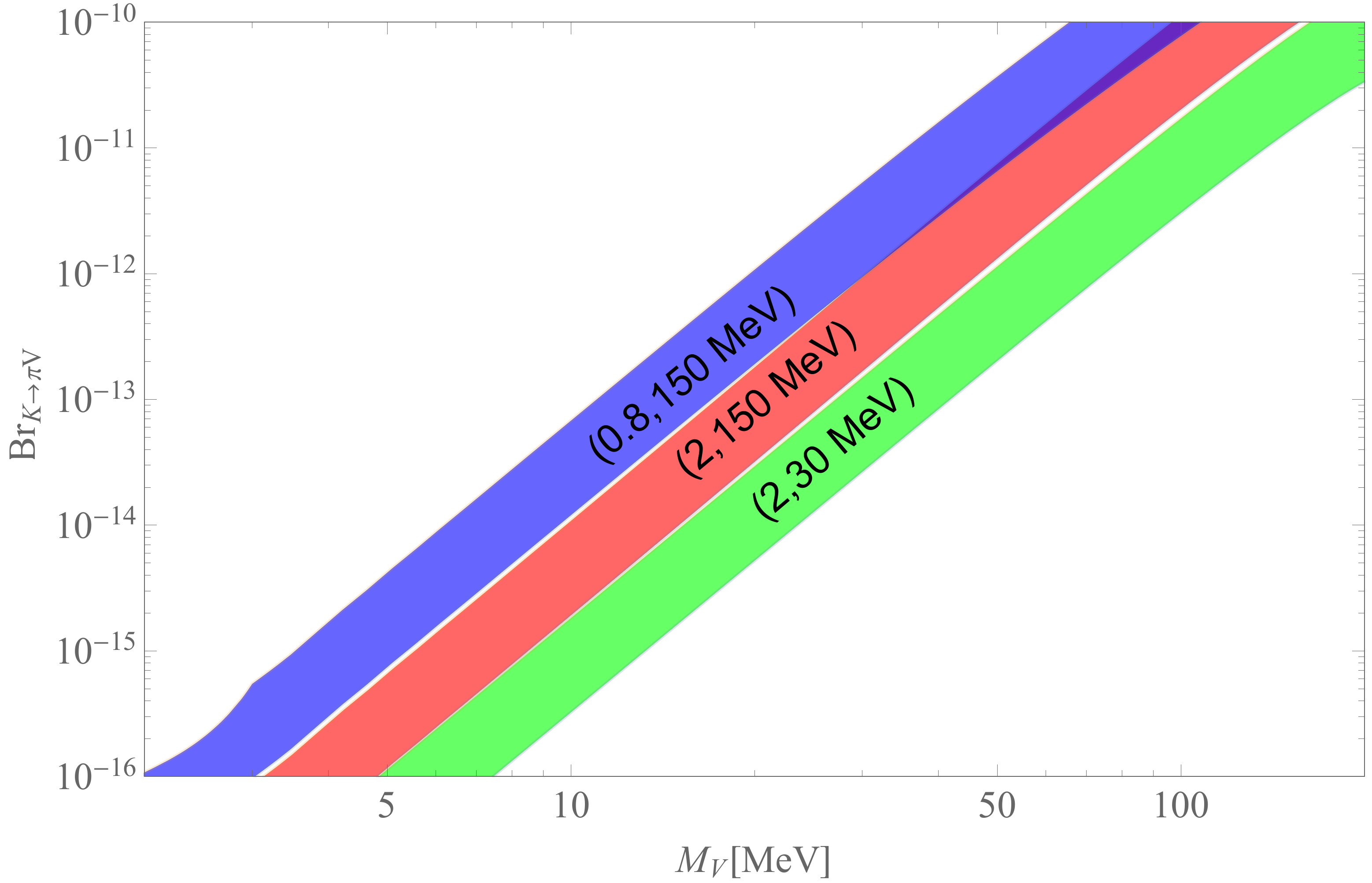}} 
	\caption[Branching ratio]{The branching ratio of $K \rightarrow \pi V$ for some specific parameters $(\lambda, m_\phi)$. The values of the coupling $\kappa$ are given by the favored region of $(g-2)_\mu$ up to $2 \sigma$. In this model $V$ decays promptly to $V \rightarrow e^+ e^-$.}\label{fig:kpiV}
\end{figure}

In the eq. (12) of \cite{Pospelov2009}, the author presented a general formula for the branching ratio of $K \to \pi V$ valid for $M_V$ below $200$ MeV and given by:

\begin{equation}
\text{Br}_{K\to \pi V} \simeq 8 \times 10^{-5} \times \kappa^2 \left(\frac{M_V}{100 \,\text{MeV}}\right)^2
\end{equation}
If we replace $\kappa^2$ by the value which explains $(g-2)_\mu$ anomaly in  eq. (\ref{gmufunction}), we can derive the branching ratio as a function of $M_V$ for a specific choice of the parameters $(\lambda, m_\phi)$. Some examples are presented in Fig. \ref{fig:kpiV}. 

The NA48/2 collaboration has commented in ref. \cite{Batley:2015lha} that sensitivity on this process is not competitive with the existing bounds. Namely, they found that 
$\kappa^2 =(0.8-1.11)\times10^{-5}$ at $90\%$ CL.  for the vector gauge boson mass in the range $2 m_e < M_V < 140$ MeV.

\subsection{$e^+ e^- \rightarrow \mu^+ \mu^- (\phi \rightarrow \mu^+ \mu^-)$ and $e^+ e^- \rightarrow \mu^+ \mu^- (V \rightarrow e^+ e^-)$} 

\begin{figure}[t]
	\centering
	\subfloat[The total cross section for $e^+ e^- \rightarrow \mu^+ \mu^- \phi, \phi \rightarrow \mu^+ \mu^-$ as function of the scalar mass $m_\phi$. The energy $\sqrt{s}$ and the constant coupling $g_\phi$ were chosen, respectively, as $10$ GeV and $10^{-3}$.]{\includegraphics[width=.45\columnwidth]{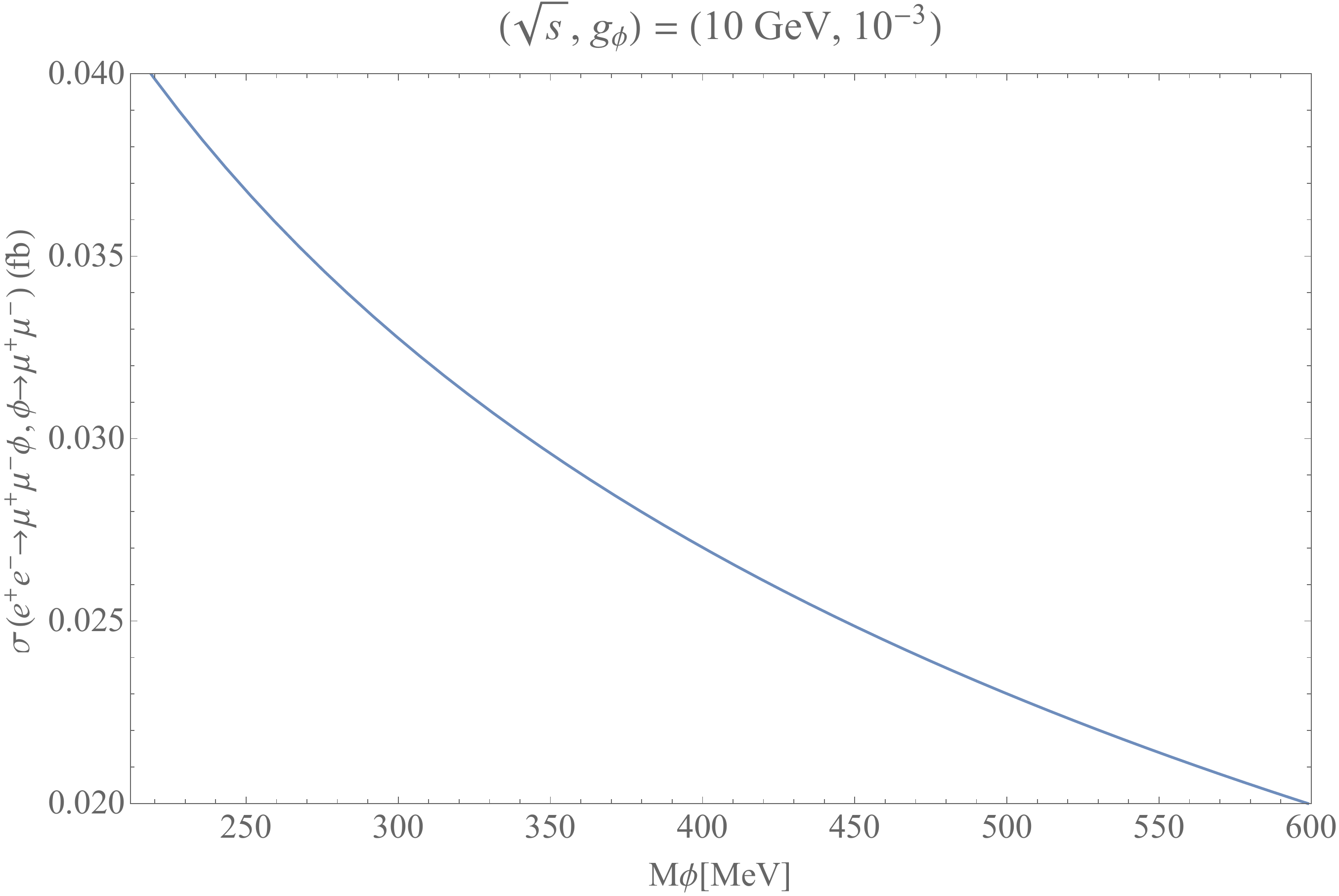}} \qquad
	\subfloat[The total cross section for $e^+ e^- \rightarrow \mu^+ \mu^- V, V \rightarrow \mu^+ \mu^-$ as function of the vector mass, $M_V$ for two values of $\lambda$.]{\includegraphics[width=.45\columnwidth]{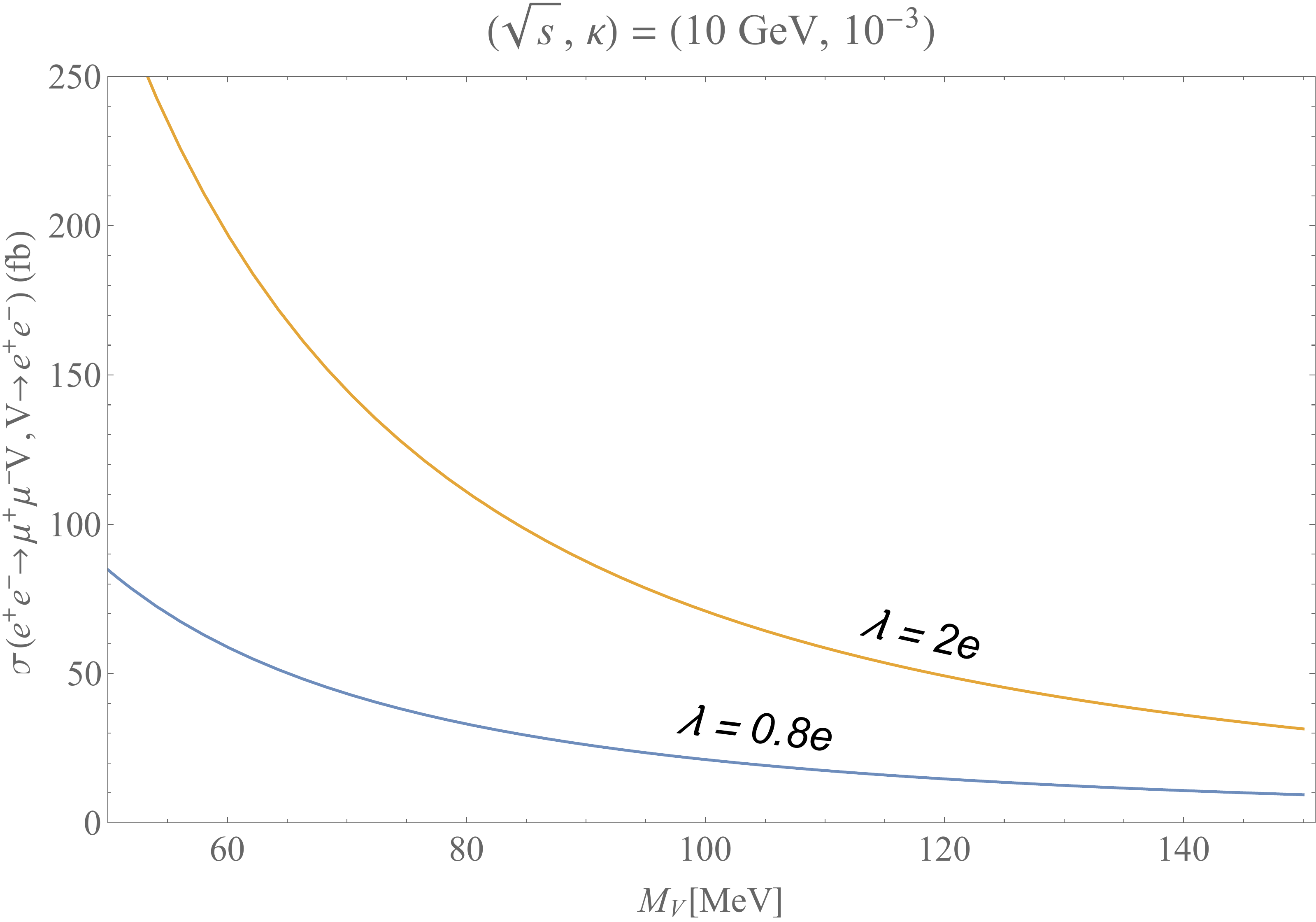}} 
	\caption[Feynman Diagrams.]{The  total cross section for $e^+ e^- \rightarrow \mu^+ \mu^- \mu^+ \mu^-$ in the framework of $\phi$ and $V$ emission. The results are complementary to the Fig. 3 of \cite{TheBABAR:2016rlg}.} 
	\label{fig:crosssec}
\end{figure}

In the work of ref. \cite{TheBABAR:2016rlg}  the search for a direct production of muonic dark forces in a model-independent method was done. The results were presented as the measured  $e^+ e^- \rightarrow \mu^+ \mu^- Z', Z' \rightarrow \mu^+ \mu^-$ cross-section being a function of the $Z'$ mass.   Within the dark  $U(1)_d$ model  we consider in this paper, the only contribution to the process $e^+ e^- \rightarrow \mu^+ \mu^- \mu^+ \mu^-$  is from the dark Higgs scalar $\phi$ in the region $M_V < 2m_\mu$ and it is presented in Fig. \ref{fig:crosssec}(a).
Nevertheless, a complementary search at  a low-mass region for $V$ will be in the process $e^+ e^- \rightarrow \mu^+ \mu^- V, V \rightarrow e^+ e^-$. The theoretical results are presented in Fig. \ref{fig:crosssec}(b) for  the  center-of-mass energy  equal to $\sqrt{10}$ GeV.

\subsection{$\pi^0  \to \gamma e^+ e^-$ and $\eta  \to \gamma e^+ e^-$  }

The dark $U(1)_d$ sector might affect  low energy observables  due to  the mixing of  the SM photon with the part of dark gauge boson.  That means in all processes where this conversion $\gamma \leftrightarrow V$ occurs,  one can search for the  dark boson. Due to the long lived vector, the use of narrow-width approximation is fully justified and we use 
$ Br ( V \to e^+ e^-) =1$. As already suggested in ref. \cite{Gardner:2015wea} the search for the presence of  dark sector in electromagnetic decays seems to be possible.
We make predictions for the decays of $P\to \gamma V \to \gamma e^+ e^-$ for $P= \pi^0$, $\eta$ by noticing that our V can have the mass in the region $50$ MeV $ <M_V< 150$ MeV  and can decay only to the electron-positron pair. In ref.  \cite{Gardner:2015wea}   $\pi^0(\eta) \to \gamma V \to \gamma e^+ e^-$ were approached by relying on the result of model independent study given in ref.  \cite{Masjuan:2012wy}. In this approach the  branching ratio of $\pi^0 \to \gamma e^+ e^-$ agrees very well with the experimental one. The decay width for 
$\pi^0 \to \gamma  V \to \gamma e^+ e^-$ with the use of the narrow-width approximation can be written as:
\begin{equation}
\Gamma (\pi^0 \to \gamma  V \to \gamma e^+ e^-) = | f(1,0, x_V)|^2 \Gamma (\pi^0 \to \gamma V) Br ( V \to e^+ e^-),\,
\label{pi0V}
\end{equation}
with $\Gamma(\pi^0 \to \gamma V) = 2 \kappa^2 (1- M_V^2/m_\pi^2)^3 \Gamma_0$, $\Gamma_0 \equiv \Gamma(\pi^0 \to \gamma \gamma)$ and 
$f(1,0,x) = 1 +a_\pi x +b_\pi x^2 + {\cal O} (x^3) $, with $a_\pi = -0.0324(12)_{stat} (19)_{sys}$, $b_\pi = 1.06(9))_{stat} (25)_{sys}\times 10^{-3}$ and $x_V= M_V^2/ m_{\pi^0}^2$, as explained in details in \cite{Gardner:2015wea}. 
For the value $M_V = 50$ MeV,  we obtained $  BR(\pi^0 \to \gamma  V \to \gamma e^+ e^-) = 1.3 \times 10^{-6} (\kappa/10^{-3})^2$, while for the 
$M_V = 100$ MeV we  calculate $ BR(\pi^0 \to \gamma  V \to \gamma e^+ e^-) = 1.9\times 10^{-7} (\kappa/10^{-3})^2$. 

For the decay of $\eta \to \gamma  V \to \gamma e^+ e^-$ following \cite{Gardner:2015wea} and using the transition form-factor from the same re\-ference 
$f_\eta (1,0,x) = 1 +b_\eta x +c_\eta x^2 + d_\eta x^3 {\cal O} (x^4) $ ($b_\eta= 0.576(11)_{stat} (4)_{sys}$,  $c_\eta=0.339(15)_{stat} (5)_{sys}$ and  $d_\eta=0.200(14)_{stat} (18)_{sys}$, and $x_V= M_V^2/ m_{\eta}^2$ \cite{Escribano:2015nra}), we obtain for
$M_V = 50$ MeV the branching ratio  $BR(\eta\to \gamma  V \to \gamma e^+ e^-) = 1.5\times 10^{-6} (\kappa/10^{-3})^2$, while for 
$M_V = 100$ MeV   we calculate $ BR(\eta \to \gamma  V \to \gamma e^+ e^-) = 1.4\times 10^{-6} (\kappa/10^{-3})^2$. 
There are few  experimental studies of  the dark matter contributions in $\pi^0 (\eta)  \to \gamma e^+ e^-$ as  described in \cite{Gardner:2015wea}  starting with the beam-dump experiments E141 \cite{Riordan:1987aw}, CHARM \cite{Gninenko:2011uv}, NA48/2 \cite{Batley:2015lha}. Also, there are plans for the future facilities as 
APEX \cite{Essig:2013lka}, HPS \cite{Boyce:2012ym}
DarkLight \cite{Freytsis:2009bh}   and LHCb\cite{Ilten:2015hya}.  The  NA48/2 experiment almost reached  sensitivity on the mixing parameter $\kappa \sim 10^{-3}$ 
\cite{Batley:2015lha} in their search for  the dark photons in $\pi^0( \eta ) \to \gamma e^+ e^-$. In \cite{Gninenko:2011uv}, based on NOMAD and PS191 it was claimed that  the bound on $BR(\pi^0  \to \gamma X \to \gamma e^+ e^-) \leq 10^{-15}$ can be reached and for the decay $BR(\eta\to  X  \gamma \to \gamma e^+ e^-) \leq 10^{-14}$ \cite{Gninenko:2012eq}. 

\subsection{$\rho \to \pi e^+ e^-$, $K^* \to K e^+ e^-$  and $\phi (1020) \to \eta e^+ e^-$}

The amplitude for the  decays $P^* \to P  V$ , $P^* = \rho^{+,0}$, $K^{* +,0}$, $\phi$ and $P= \pi^{+,0}$, $K^{+,0}$ and $\eta$ can be written as:
\begin{equation}
{\cal M}(P^* (p_{P^*},  \epsilon_{P^*}) \to P (p_P )  V (p_V,\epsilon_V )= \kappa \, g_{P^* PV} \epsilon_{\mu \nu \alpha \beta} p_{P^*}^\mu \epsilon_ {P^*}^\nu   p_{V}^\alpha \epsilon_V ^\beta , 
\label{radV}
\end{equation}
with $p_{P}$, $p_{P^*} $ and  $P_V$ being the momenta of the corresponding mesons and $\epsilon_{P^*} $ and $\epsilon_V $ being polarization vectors of $P^*$ and $V$, respectively. 
In order to determine decay widths, we assume that to a good approximation $g_{P^* PV}\simeq g_{P^* P\gamma}$. 
We expect that this approximation is satisfied as long as the dark vector boson mass is relatively small. There are numerous attempts within lattice QCD community to calculate $\rho \pi \gamma^*$ form-factors  \cite{Briceno:2016kkp} which will help in more precise studies of $g_{P^* PV}$. The transition coefficient $g_{P^* P\gamma}$ can be extracted from the decay width  for $P^* \to P  \gamma$. Knowing that $\Gamma (P^* \to P  \gamma)= | g_{P^* P\gamma}|^2(m_{P^*}^2 -m_P^2)^3/(96 \pi  m_{P^*}^3)$, one can determine $g_{P^* P\gamma}$. It was found by  the authors of   \cite{YamagataSekihara:2010ip} that 
  $g_{\rho^+ \pi^+ \gamma} =2.19\times 10^{-4}$ MeV$^{-1}$,  $g_{\rho^0 \pi^0\gamma} =2.52\times 10^{-4}$ MeV
  $^{-1}$, $g_{K^{*+} K^+\gamma} =2.53\times 10^{-4}$ MeV$^{-1}$,
  $g_{K^{*0} K^0\gamma} =2.19\times 10^{-4}$ MeV$^{-1}$ and we obtain the value $g_{\Phi(1020) \eta \gamma} = 1.26 \times 10^{-4}$ MeV$^{-1}$, using data given in 
 PDG \cite{Agashe:2014kda}. The decay width for $P^* \to P  V$  can be written as:
 \begin{equation}
 \Gamma (P^* \to P  V) = \frac{ |\kappa \,g_{P^* PV}|^2}{96 \pi} \frac{\lambda (m_{P^*}^2, m_P^2, M_V^2)^{3/2}}{m_{P^*}^3}\, ,
 \label{width-vector1}
 \end{equation}
and finally:

\begin{equation}
 \Gamma (P^* \to P  V \to P e^+ e^-) =  \Gamma (P^* \to P  V)  BR(V \to e^+ e^-).
 \label{width-vector}
 \end{equation}

 \begin{table}[h!]
  \centering
  \caption{Predicted branching ratios for $BR(P^* \to P  V \to P e^+ e^-) $ for the dark gauge boson mass $M_V =50, 100$ MeV and $\kappa=0.001$. For other values of the photon-V mixing parameter $\kappa$ one should rescale these results  by $ (\kappa/10^{-3})^2$. }
  \label{tab:table1}
  \begin{tabular}{|l||c|c|}
  \hline
$ P^*, P,  V$    &  $M_V =50\, MeV$ & $M_V =100\, MeV$ \\
    \hline \hline
  $\rho^0, \pi^0 ,V$    & $ 6.3 \times 10^{-10}$  &  $ 6.1 \times 10^{-10}$ \\
    $\rho^+, \pi^+ ,V$    & $ 4.8 \times 10^{-10}$  &  $ 4.6\times 10^{-10}$ \\
     $K^{*0}, K^0 ,V$    & $ 7.6 \times 10^{-10}$  &  $ 7.0 \times 10^{-10}$ \\
        $K^{*+} ,K^+ ,V$    & $ 1.0 \times 10^{-9}$  &  $ 9.5\times 10^{-10}$ \\
         $\Phi(1020) ,\eta ,V$    & $ 9.1 \times 10^{-10}$  &  $  8.9\times 10^{-10}$ \\ 
          \hline
  \end{tabular}
\end{table}
There are a number of planned experimental searches in which above-mentioned processes might be relevant as
APEX \cite{Essig:2013lka}, HPS \cite{Boyce:2012ym}
DarkLight \cite{Freytsis:2009bh}   and LHCb\cite{Ilten:2015hya}.
The KLOE-2 experiment   has already searched for the dark photon contribution in $\phi \to \eta V \to V e^+ e^-$ decay \cite{Archilli:2011zc} not finding any bump in the differential distribution. 

\section{Summary}

The $U(1)_d$ gauge model of eq.(\ref{e1}) introduces a new dark gauge boson and a dark Higgs.  The model was first proposed to explain the  proton charge radius discrepancy as well as the $(g-2)_\mu$ anomaly.

We have explored the phenomenology of these two dark bosons - the dark Higgs $\phi$ and the vector $V$ - through a set of low energy processes, focusing on the parameter space $(M_V, \kappa)$. In our approach the muon magnetic moment receives the contribution of both particles and the $\phi$ mass provides an additional freedom to adjust the allowed band of $(g-2)_\mu$ within $2 \sigma$. We find out that $V $ has to decay to $e^+ e^-$ in order to explain $K\to\mu X$, where $X$ refers to a missing energy, implying that in this process any signature of the dark Higgs could be detected. Further, we  concluded that the allowed band for the proton radius anomaly is strongly constrained by a set of well-established bounds, namely  the bounds from $(g-2)_e$ and $\tau \to \nu_\tau \mu \nu_\mu e^+ e^-$ decay. 
This feature, for instance, will enable different ranges for $M_V$ which were at first excluded in the context of a generic vector coupling by the BaBar searches  \cite{TheBABAR:2016rlg}, as presented in Fig. \ref{fig:PS}(c).    
The bounds from $K \to \mu \nu_\mu  e^+ e^-$, $K \to \mu \nu_\mu  \mu^+ \mu^-$, $\tau \to \nu_\tau \mu \nu_\mu e^+ e^-$ when combined with above-mentioned bounds allow the mass of $M_V$ to be in the region around $50 \, \text{MeV} < M_V < 150 \, \text{MeV}$ with the parameter $\kappa \sim 10^{-3}$, while the mass of the dark Higgs can be from few MeV till $\sim 200$ MeV, for a particular choice of the remaining parameter $\lambda$. We finally mention that the bound from $K\to \mu \nu_\mu e^+ e^-$ leads to constraints as strong as the experimentally achieved by the BaBar and NA48/2 collaborations on the correspondent sector. 

We have also presented a set of predictions. The very small branching ratios of the processes   $K \to \pi V, V \to e^+ e^-$, as it was pointed in \cite{Pospelov2009},  
makes the search for the dark gauge boson rather difficult. The electromagnetic decays of $\pi^0  \to \gamma e^+ e^-$ and $\eta  \to \gamma e^+ e^-$ , $\rho \to \pi e^+ e^-$, $K^* \to K e^+ e^-$  and $\phi (1020) \to \eta e^+ e^-$  might also proceed through the dark gauge boson. Some of these processes are already subjects of experimental studies. The small mixing parameter between the photon and dark-photon suppresses the branching ratios for these processes, but hopefully future experiments for the dark matter search would shed more light on dark bosons at low energies.

\section{Acknowledgments}
F.C.C. appreciates hospitality during his visit to J. Stefan Institute and would like to thank D. Faroughy for important discussions. The work of FCC was supported in part by JSI and the National Counsel of Technological and Scientific Development, CNPq-Brazil. SF acknowledge support of  the Slovenian Research Agency.



\newpage
\begin{thebibliography}{60}
	\expandafter\ifx\csname natexlab\endcsname\relax\def\natexlab#1{#1}\fi
	\expandafter\ifx\csname bibnamefont\endcsname\relax
	\def\bibnamefont#1{#1}\fi
	\expandafter\ifx\csname bibfnamefont\endcsname\relax
	\def\bibfnamefont#1{#1}\fi
	\expandafter\ifx\csname citenamefont\endcsname\relax
	\def\citenamefont#1{#1}\fi
	\expandafter\ifx\csname url\endcsname\relax
	\def\url#1{\texttt{#1}}\fi
	\expandafter\ifx\csname urlprefix\endcsname\relax\def\urlprefix{URL }\fi
	\providecommand{\bibinfo}[2]{#2}
	\providecommand{\eprint}[2][]{\url{#2}}
	
	\bibitem[{\citenamefont{Davoudiasl et~al.}(2014)\citenamefont{Davoudiasl, Lee,
			and Marciano}}]{Davoudiasl:2014kua}
	\bibinfo{author}{\bibfnamefont{H.}~\bibnamefont{Davoudiasl}},
	\bibinfo{author}{\bibfnamefont{H.-S.} \bibnamefont{Lee}}, \bibnamefont{and}
	\bibinfo{author}{\bibfnamefont{W.~J.} \bibnamefont{Marciano}},
	\bibinfo{journal}{Phys. Rev.} \textbf{\bibinfo{volume}{D89}},
	\bibinfo{pages}{095006} (\bibinfo{year}{2014}), \eprint{1402.3620}.
	
	\bibitem[{\citenamefont{Batell et~al.}(2009)\citenamefont{Batell, Pospelov, and
			Ritz}}]{Batell:2009yf}
	\bibinfo{author}{\bibfnamefont{B.}~\bibnamefont{Batell}},
	\bibinfo{author}{\bibfnamefont{M.}~\bibnamefont{Pospelov}}, \bibnamefont{and}
	\bibinfo{author}{\bibfnamefont{A.}~\bibnamefont{Ritz}},
	\bibinfo{journal}{Phys. Rev.} \textbf{\bibinfo{volume}{D79}},
	\bibinfo{pages}{115008} (\bibinfo{year}{2009}), \eprint{0903.0363}.
	
	\bibitem[{\citenamefont{Batell et~al.}(2011)\citenamefont{Batell, McKeen, and
			Pospelov}}]{Batell:2011qq}
	\bibinfo{author}{\bibfnamefont{B.}~\bibnamefont{Batell}},
	\bibinfo{author}{\bibfnamefont{D.}~\bibnamefont{McKeen}}, \bibnamefont{and}
	\bibinfo{author}{\bibfnamefont{M.}~\bibnamefont{Pospelov}},
	\bibinfo{journal}{Phys. Rev. Lett.} \textbf{\bibinfo{volume}{107}},
	\bibinfo{pages}{011803} (\bibinfo{year}{2011}), \eprint{1103.0721}.
	
	\bibitem[{\citenamefont{Fayet}(2007)}]{Fayet:2007ua}
	\bibinfo{author}{\bibfnamefont{P.}~\bibnamefont{Fayet}},
	\bibinfo{journal}{Phys. Rev.} \textbf{\bibinfo{volume}{D75}},
	\bibinfo{pages}{115017} (\bibinfo{year}{2007}), \eprint{hep-ph/0702176}.
	
	\bibitem[{\citenamefont{Carlson and Rislow}(2014)}]{Carlson:2013mya}
	\bibinfo{author}{\bibfnamefont{C.~E.} \bibnamefont{Carlson}} \bibnamefont{and}
	\bibinfo{author}{\bibfnamefont{B.~C.} \bibnamefont{Rislow}},
	\bibinfo{journal}{Phys. Rev.} \textbf{\bibinfo{volume}{D89}},
	\bibinfo{pages}{035003} (\bibinfo{year}{2014}), \eprint{1310.2786}.
	
	\bibitem[{\citenamefont{Davoudiasl
			et~al.}(2012{\natexlab{a}})\citenamefont{Davoudiasl, Lee, and
			Marciano}}]{Davoudiasl:2012qa}
	\bibinfo{author}{\bibfnamefont{H.}~\bibnamefont{Davoudiasl}},
	\bibinfo{author}{\bibfnamefont{H.-S.} \bibnamefont{Lee}}, \bibnamefont{and}
	\bibinfo{author}{\bibfnamefont{W.~J.} \bibnamefont{Marciano}},
	\bibinfo{journal}{Phys. Rev. Lett.} \textbf{\bibinfo{volume}{109}},
	\bibinfo{pages}{031802} (\bibinfo{year}{2012}{\natexlab{a}}),
	\eprint{1205.2709}.
	
	\bibitem[{\citenamefont{Lee}(2014)}]{Lee:2014tba}
	\bibinfo{author}{\bibfnamefont{H.-S.} \bibnamefont{Lee}},
	\bibinfo{journal}{Phys. Rev.} \textbf{\bibinfo{volume}{D90}},
	\bibinfo{pages}{091702} (\bibinfo{year}{2014}), \eprint{1408.4256}.
	
	\bibitem[{\citenamefont{Karshenboim et~al.}(2014)\citenamefont{Karshenboim,
			McKeen, and Pospelov}}]{Karshenboim:2014tka}
	\bibinfo{author}{\bibfnamefont{S.~G.} \bibnamefont{Karshenboim}},
	\bibinfo{author}{\bibfnamefont{D.}~\bibnamefont{McKeen}}, \bibnamefont{and}
	\bibinfo{author}{\bibfnamefont{M.}~\bibnamefont{Pospelov}},
	\bibinfo{journal}{Phys. Rev.} \textbf{\bibinfo{volume}{D90}},
	\bibinfo{pages}{073004} (\bibinfo{year}{2014}), \bibinfo{note}{[Addendum:
		Phys. Rev.D90,no.7,079905(2014)]}, \eprint{1401.6154}.
	
	\bibitem[{\citenamefont{Batell et~al.}(2016)\citenamefont{Batell, Lange,
			McKeen, Pospelov, and Ritz}}]{Batell:2016ove}
	\bibinfo{author}{\bibfnamefont{B.}~\bibnamefont{Batell}},
	\bibinfo{author}{\bibfnamefont{N.}~\bibnamefont{Lange}},
	\bibinfo{author}{\bibfnamefont{D.}~\bibnamefont{McKeen}},
	\bibinfo{author}{\bibfnamefont{M.}~\bibnamefont{Pospelov}}, \bibnamefont{and}
	\bibinfo{author}{\bibfnamefont{A.}~\bibnamefont{Ritz}}
	(\bibinfo{year}{2016}), \eprint{1606.04943}.
	
	\bibitem[{\citenamefont{Barger et~al.}(2011)\citenamefont{Barger, Chiang,
			Keung, and Marfatia}}]{Barger:2010aj}
	\bibinfo{author}{\bibfnamefont{V.}~\bibnamefont{Barger}},
	\bibinfo{author}{\bibfnamefont{C.-W.} \bibnamefont{Chiang}},
	\bibinfo{author}{\bibfnamefont{W.-Y.} \bibnamefont{Keung}}, \bibnamefont{and}
	\bibinfo{author}{\bibfnamefont{D.}~\bibnamefont{Marfatia}},
	\bibinfo{journal}{Phys. Rev. Lett.} \textbf{\bibinfo{volume}{106}},
	\bibinfo{pages}{153001} (\bibinfo{year}{2011}), \eprint{1011.3519}.
	
	\bibitem[{\citenamefont{Carlson}(2015)}]{Carlson:2015jba}
	\bibinfo{author}{\bibfnamefont{C.~E.} \bibnamefont{Carlson}},
	\bibinfo{journal}{Prog. Part. Nucl. Phys.} \textbf{\bibinfo{volume}{82}},
	\bibinfo{pages}{59} (\bibinfo{year}{2015}), \eprint{1502.05314}.
	
	\bibitem[{\citenamefont{Carlson and Rislow}(2012)}]{Carlson:2012pc}
	\bibinfo{author}{\bibfnamefont{C.~E.} \bibnamefont{Carlson}} \bibnamefont{and}
	\bibinfo{author}{\bibfnamefont{B.~C.} \bibnamefont{Rislow}},
	\bibinfo{journal}{Phys. Rev.} \textbf{\bibinfo{volume}{D86}},
	\bibinfo{pages}{035013} (\bibinfo{year}{2012}), \eprint{1206.3587}.
	
	\bibitem[{\citenamefont{Arkani-Hamed et~al.}(2009)\citenamefont{Arkani-Hamed,
			Finkbeiner, Slatyer, and Weiner}}]{ArkaniHamed:2008qn}
	\bibinfo{author}{\bibfnamefont{N.}~\bibnamefont{Arkani-Hamed}},
	\bibinfo{author}{\bibfnamefont{D.~P.} \bibnamefont{Finkbeiner}},
	\bibinfo{author}{\bibfnamefont{T.~R.} \bibnamefont{Slatyer}},
	\bibnamefont{and} \bibinfo{author}{\bibfnamefont{N.}~\bibnamefont{Weiner}},
	\bibinfo{journal}{Phys. Rev.} \textbf{\bibinfo{volume}{D79}},
	\bibinfo{pages}{015014} (\bibinfo{year}{2009}), \eprint{0810.0713}.
	
	\bibitem[{\citenamefont{Boehm et~al.}(2004)\citenamefont{Boehm, Hooper, Silk,
			Casse, and Paul}}]{Boehm:2003bt}
	\bibinfo{author}{\bibfnamefont{C.}~\bibnamefont{Boehm}},
	\bibinfo{author}{\bibfnamefont{D.}~\bibnamefont{Hooper}},
	\bibinfo{author}{\bibfnamefont{J.}~\bibnamefont{Silk}},
	\bibinfo{author}{\bibfnamefont{M.}~\bibnamefont{Casse}}, \bibnamefont{and}
	\bibinfo{author}{\bibfnamefont{J.}~\bibnamefont{Paul}},
	\bibinfo{journal}{Phys. Rev. Lett.} \textbf{\bibinfo{volume}{92}},
	\bibinfo{pages}{101301} (\bibinfo{year}{2004}), \eprint{astro-ph/0309686}.
	
	\bibitem[{\citenamefont{Essig et~al.}(2013)}]{Essig:2013lka}
	\bibinfo{author}{\bibfnamefont{R.}~\bibnamefont{Essig}} \bibnamefont{et~al.},
	in \emph{\bibinfo{booktitle}{{Proceedings, Community Summer Study 2013:
				Snowmass on the Mississippi (CSS2013): Minneapolis, MN, USA, July 29-August
				6, 2013}}} (\bibinfo{year}{2013}), \eprint{1311.0029},
	\urlprefix\url{https://inspirehep.net/record/1263039/files/arXiv:1311.0029.pdf}.
	
	\bibitem[{\citenamefont{Lees et~al.}(2016)}]{TheBABAR:2016rlg}
	\bibinfo{author}{\bibfnamefont{J.~P.} \bibnamefont{Lees}} \bibnamefont{et~al.}
	(\bibinfo{collaboration}{BaBar}) (\bibinfo{year}{2016}), \eprint{1606.03501}.
	
	\bibitem[{\citenamefont{Lees et~al.}(2014)}]{Lees:2014xha}
	\bibinfo{author}{\bibfnamefont{J.~P.} \bibnamefont{Lees}} \bibnamefont{et~al.}
	(\bibinfo{collaboration}{BaBar}), \bibinfo{journal}{Phys. Rev. Lett.}
	\textbf{\bibinfo{volume}{113}}, \bibinfo{pages}{201801}
	(\bibinfo{year}{2014}), \eprint{1406.2980}.
	
	\bibitem[{\citenamefont{Cable et~al.}(1973)\citenamefont{Cable, Hildebrand,
			Pang, and Stiening}}]{Cable:1974nx}
	\bibinfo{author}{\bibfnamefont{G.~D.} \bibnamefont{Cable}},
	\bibinfo{author}{\bibfnamefont{R.~H.} \bibnamefont{Hildebrand}},
	\bibinfo{author}{\bibfnamefont{C.~Y.} \bibnamefont{Pang}}, \bibnamefont{and}
	\bibinfo{author}{\bibfnamefont{R.}~\bibnamefont{Stiening}},
	\bibinfo{journal}{Phys. Rev.} \textbf{\bibinfo{volume}{D8}},
	\bibinfo{pages}{3807} (\bibinfo{year}{1973}).
	
	\bibitem[{\citenamefont{Barger et~al.}(2012)\citenamefont{Barger, Chiang,
			Keung, and Marfatia}}]{Barger:2011mt}
	\bibinfo{author}{\bibfnamefont{V.}~\bibnamefont{Barger}},
	\bibinfo{author}{\bibfnamefont{C.-W.} \bibnamefont{Chiang}},
	\bibinfo{author}{\bibfnamefont{W.-Y.} \bibnamefont{Keung}}, \bibnamefont{and}
	\bibinfo{author}{\bibfnamefont{D.}~\bibnamefont{Marfatia}},
	\bibinfo{journal}{Phys. Rev. Lett.} \textbf{\bibinfo{volume}{108}},
	\bibinfo{pages}{081802} (\bibinfo{year}{2012}), \eprint{1109.6652}.
	
	\bibitem[{\citenamefont{Goertz et~al.}(2016)\citenamefont{Goertz, Kamenik,
			Katz, and Nardecchia}}]{Goertz:2015nkp}
	\bibinfo{author}{\bibfnamefont{F.}~\bibnamefont{Goertz}},
	\bibinfo{author}{\bibfnamefont{J.~F.} \bibnamefont{Kamenik}},
	\bibinfo{author}{\bibfnamefont{A.}~\bibnamefont{Katz}}, \bibnamefont{and}
	\bibinfo{author}{\bibfnamefont{M.}~\bibnamefont{Nardecchia}},
	\bibinfo{journal}{JHEP} \textbf{\bibinfo{volume}{05}}, \bibinfo{pages}{187}
	(\bibinfo{year}{2016}), \eprint{1512.08500}.
	
	\bibitem[{\citenamefont{Fileviez~Perez and Wise}(2010)}]{FileviezPerez:2010gw}
	\bibinfo{author}{\bibfnamefont{P.}~\bibnamefont{Fileviez~Perez}}
	\bibnamefont{and} \bibinfo{author}{\bibfnamefont{M.~B.} \bibnamefont{Wise}},
	\bibinfo{journal}{Phys. Rev.} \textbf{\bibinfo{volume}{D82}},
	\bibinfo{pages}{011901} (\bibinfo{year}{2010}), \bibinfo{note}{[Erratum:
		Phys. Rev.D82,079901(2010)]}, \eprint{1002.1754}.
	
	\bibitem[{\citenamefont{Chao}(2011)}]{Chao:2010mp}
	\bibinfo{author}{\bibfnamefont{W.}~\bibnamefont{Chao}}, \bibinfo{journal}{Phys.
		Lett.} \textbf{\bibinfo{volume}{B695}}, \bibinfo{pages}{157}
	(\bibinfo{year}{2011}), \eprint{1005.1024}.
	
	\bibitem[{\citenamefont{Ko and Omura}(2011)}]{Ko:2010at}
	\bibinfo{author}{\bibfnamefont{P.}~\bibnamefont{Ko}} \bibnamefont{and}
	\bibinfo{author}{\bibfnamefont{Y.}~\bibnamefont{Omura}},
	\bibinfo{journal}{Phys. Lett.} \textbf{\bibinfo{volume}{B701}},
	\bibinfo{pages}{363} (\bibinfo{year}{2011}), \eprint{1012.4679}.
	
	\bibitem[{\citenamefont{Duerr et~al.}(2013)\citenamefont{Duerr, Fileviez~Perez,
			and Wise}}]{Duerr:2013dza}
	\bibinfo{author}{\bibfnamefont{M.}~\bibnamefont{Duerr}},
	\bibinfo{author}{\bibfnamefont{P.}~\bibnamefont{Fileviez~Perez}},
	\bibnamefont{and} \bibinfo{author}{\bibfnamefont{M.~B.} \bibnamefont{Wise}},
	\bibinfo{journal}{Phys. Rev. Lett.} \textbf{\bibinfo{volume}{110}},
	\bibinfo{pages}{231801} (\bibinfo{year}{2013}), \eprint{1304.0576}.
	
	\bibitem[{\citenamefont{Schwaller et~al.}(2013)\citenamefont{Schwaller, Tait,
			and Vega-Morales}}]{Schwaller:2013hqa}
	\bibinfo{author}{\bibfnamefont{P.}~\bibnamefont{Schwaller}},
	\bibinfo{author}{\bibfnamefont{T.~M.~P.} \bibnamefont{Tait}},
	\bibnamefont{and}
	\bibinfo{author}{\bibfnamefont{R.}~\bibnamefont{Vega-Morales}},
	\bibinfo{journal}{Phys. Rev.} \textbf{\bibinfo{volume}{D88}},
	\bibinfo{pages}{035001} (\bibinfo{year}{2013}), \eprint{1305.1108}.
	
	\bibitem[{\citenamefont{Chen et~al.}(2016)\citenamefont{Chen, Davoudiasl,
			Marciano, and Zhang}}]{Chen:2015vqy}
	\bibinfo{author}{\bibfnamefont{C.-Y.} \bibnamefont{Chen}},
	\bibinfo{author}{\bibfnamefont{H.}~\bibnamefont{Davoudiasl}},
	\bibinfo{author}{\bibfnamefont{W.~J.} \bibnamefont{Marciano}},
	\bibnamefont{and} \bibinfo{author}{\bibfnamefont{C.}~\bibnamefont{Zhang}},
	\bibinfo{journal}{Phys. Rev.} \textbf{\bibinfo{volume}{D93}},
	\bibinfo{pages}{035006} (\bibinfo{year}{2016}), \eprint{1511.04715}.
	
	\bibitem[{\citenamefont{Antognini et~al.}(2013)\citenamefont{Antognini,
			Kottmann, Biraben, Indelicato, Nez, and Pohl}}]{Antognini:2013jkc}
	\bibinfo{author}{\bibfnamefont{A.}~\bibnamefont{Antognini}},
	\bibinfo{author}{\bibfnamefont{F.}~\bibnamefont{Kottmann}},
	\bibinfo{author}{\bibfnamefont{F.}~\bibnamefont{Biraben}},
	\bibinfo{author}{\bibfnamefont{P.}~\bibnamefont{Indelicato}},
	\bibinfo{author}{\bibfnamefont{F.}~\bibnamefont{Nez}}, \bibnamefont{and}
	\bibinfo{author}{\bibfnamefont{R.}~\bibnamefont{Pohl}},
	\bibinfo{journal}{Annals Phys.} \textbf{\bibinfo{volume}{331}},
	\bibinfo{pages}{127} (\bibinfo{year}{2013}), \eprint{1208.2637}.
	
	\bibitem[{\citenamefont{Pohl et~al.}(2010)}]{Pohl:2010zza}
	\bibinfo{author}{\bibfnamefont{R.}~\bibnamefont{Pohl}} \bibnamefont{et~al.},
	\bibinfo{journal}{Nature} \textbf{\bibinfo{volume}{466}},
	\bibinfo{pages}{213} (\bibinfo{year}{2010}).
	
	\bibitem[{\citenamefont{Olive et~al.}(2014)}]{Agashe:2014kda}
	\bibinfo{author}{\bibfnamefont{K.~A.} \bibnamefont{Olive}} \bibnamefont{et~al.}
	(\bibinfo{collaboration}{Particle Data Group}), \bibinfo{journal}{Chin.
		Phys.} \textbf{\bibinfo{volume}{C38}}, \bibinfo{pages}{090001}
	(\bibinfo{year}{2014}).
	
	\bibitem[{\citenamefont{Leveille and Weiler}(1979)}]{Leveille:1978px}
	\bibinfo{author}{\bibfnamefont{J.~P.} \bibnamefont{Leveille}} \bibnamefont{and}
	\bibinfo{author}{\bibfnamefont{T.~J.} \bibnamefont{Weiler}},
	\bibinfo{journal}{Nucl. Phys.} \textbf{\bibinfo{volume}{B147}},
	\bibinfo{pages}{147} (\bibinfo{year}{1979}).
	
	\bibitem[{\citenamefont{McKeen}(2011)}]{McKeen:2009ny}
	\bibinfo{author}{\bibfnamefont{D.}~\bibnamefont{McKeen}},
	\bibinfo{journal}{Annals Phys.} \textbf{\bibinfo{volume}{326}},
	\bibinfo{pages}{1501} (\bibinfo{year}{2011}), \eprint{0912.1076}.
	
	\bibitem[{\citenamefont{Bijnens et~al.}(1993)\citenamefont{Bijnens, Ecker, and
			Gasser}}]{Bijnens:1992en}
	\bibinfo{author}{\bibfnamefont{J.}~\bibnamefont{Bijnens}},
	\bibinfo{author}{\bibfnamefont{G.}~\bibnamefont{Ecker}}, \bibnamefont{and}
	\bibinfo{author}{\bibfnamefont{J.}~\bibnamefont{Gasser}},
	\bibinfo{journal}{Nucl. Phys.} \textbf{\bibinfo{volume}{B396}},
	\bibinfo{pages}{81} (\bibinfo{year}{1993}), \eprint{hep-ph/9209261}.
	
	\bibitem[{\citenamefont{Cirigliano and Rosell}(2007)}]{Cirigliano:2007xi}
	\bibinfo{author}{\bibfnamefont{V.}~\bibnamefont{Cirigliano}} \bibnamefont{and}
	\bibinfo{author}{\bibfnamefont{I.}~\bibnamefont{Rosell}},
	\bibinfo{journal}{Phys. Rev. Lett.} \textbf{\bibinfo{volume}{99}},
	\bibinfo{pages}{231801} (\bibinfo{year}{2007}), \eprint{0707.3439}.
	
	\bibitem[{\citenamefont{Poblaguev et~al.}(2002)}]{Poblaguev:2002ug}
	\bibinfo{author}{\bibfnamefont{A.~A.} \bibnamefont{Poblaguev}}
	\bibnamefont{et~al.}, \bibinfo{journal}{Phys. Rev. Lett.}
	\textbf{\bibinfo{volume}{89}}, \bibinfo{pages}{061803}
	(\bibinfo{year}{2002}), \eprint{hep-ex/0204006}.
	
	\bibitem[{\citenamefont{Aoyama et~al.}(2012)\citenamefont{Aoyama, Hayakawa,
			Kinoshita, and Nio}}]{Aoyama:2012wj}
	\bibinfo{author}{\bibfnamefont{T.}~\bibnamefont{Aoyama}},
	\bibinfo{author}{\bibfnamefont{M.}~\bibnamefont{Hayakawa}},
	\bibinfo{author}{\bibfnamefont{T.}~\bibnamefont{Kinoshita}},
	\bibnamefont{and} \bibinfo{author}{\bibfnamefont{M.}~\bibnamefont{Nio}},
	\bibinfo{journal}{Phys. Rev. Lett.} \textbf{\bibinfo{volume}{109}},
	\bibinfo{pages}{111807} (\bibinfo{year}{2012}), \eprint{1205.5368}.
	
	\bibitem[{\citenamefont{Pospelov}(2009)}]{Pospelov2009}
	\bibinfo{author}{\bibfnamefont{M.}~\bibnamefont{Pospelov}},
	\bibinfo{journal}{Phys. Rev. D} \textbf{\bibinfo{volume}{80}},
	\bibinfo{pages}{095002} (\bibinfo{year}{2009}),
	\urlprefix\url{http://link.aps.org/doi/10.1103/PhysRevD.80.095002}.
	
	\bibitem[{\citenamefont{Babusci et~al.}(2014)}]{Babusci:2014sta}
	\bibinfo{author}{\bibfnamefont{D.}~\bibnamefont{Babusci}} \bibnamefont{et~al.}
	(\bibinfo{collaboration}{KLOE-2}), \bibinfo{journal}{Phys. Lett.}
	\textbf{\bibinfo{volume}{B736}}, \bibinfo{pages}{459} (\bibinfo{year}{2014}),
	\eprint{1404.7772}.
	
	\bibitem[{\citenamefont{Merkel et~al.}(2014)}]{Merkel:2014avp}
	\bibinfo{author}{\bibfnamefont{H.}~\bibnamefont{Merkel}} \bibnamefont{et~al.},
	\bibinfo{journal}{Phys. Rev. Lett.} \textbf{\bibinfo{volume}{112}},
	\bibinfo{pages}{221802} (\bibinfo{year}{2014}), \eprint{1404.5502}.
	
	\bibitem[{\citenamefont{Perez~del Rio}(2016)}]{delRio:2016anz}
	\bibinfo{author}{\bibfnamefont{E.}~\bibnamefont{Perez~del Rio}}
	(\bibinfo{collaboration}{KLOE-2}) (\bibinfo{year}{2016}),
	\eprint{1602.00492},
	\urlprefix\url{https://inspirehep.net/record/1418817/files/arXiv:1602.00492.pdf}.
	
	\bibitem[{\citenamefont{Batley et~al.}(2015)}]{Batley:2015lha}
	\bibinfo{author}{\bibfnamefont{J.~R.} \bibnamefont{Batley}}
	\bibnamefont{et~al.} (\bibinfo{collaboration}{NA48/2}),
	\bibinfo{journal}{Phys. Lett.} \textbf{\bibinfo{volume}{B746}},
	\bibinfo{pages}{178} (\bibinfo{year}{2015}), \eprint{1504.00607}.
	
	\bibitem[{\citenamefont{Anastasi et~al.}(2015)}]{Anastasi:2015qla}
	\bibinfo{author}{\bibfnamefont{A.}~\bibnamefont{Anastasi}}
	\bibnamefont{et~al.}, \bibinfo{journal}{Phys. Lett.}
	\textbf{\bibinfo{volume}{B750}}, \bibinfo{pages}{633} (\bibinfo{year}{2015}),
	\eprint{1509.00740}.
	
	\bibitem[{\citenamefont{Anastasi et~al.}(2016)}]{::2016lwm}
	\bibinfo{author}{\bibfnamefont{A.}~\bibnamefont{Anastasi}} \bibnamefont{et~al.}
	(\bibinfo{collaboration}{KLOE-2}), \bibinfo{journal}{Phys. Lett.}
	\textbf{\bibinfo{volume}{B757}}, \bibinfo{pages}{356} (\bibinfo{year}{2016}),
	\eprint{1603.06086}.
	
	\bibitem[{\citenamefont{Archilli et~al.}(2012)}]{Archilli:2011zc}
	\bibinfo{author}{\bibfnamefont{F.}~\bibnamefont{Archilli}} \bibnamefont{et~al.}
	(\bibinfo{collaboration}{KLOE-2}), \bibinfo{journal}{Phys. Lett.}
	\textbf{\bibinfo{volume}{B706}}, \bibinfo{pages}{251} (\bibinfo{year}{2012}),
	\eprint{1110.0411}.
	
	\bibitem[{\citenamefont{Bjorken et~al.}(1988)\citenamefont{Bjorken, Ecklund,
			Nelson, Abashian, Church, Lu, Mo, Nunamaker, and Rassmann}}]{Bjorken:1988as}
	\bibinfo{author}{\bibfnamefont{J.~D.} \bibnamefont{Bjorken}},
	\bibinfo{author}{\bibfnamefont{S.}~\bibnamefont{Ecklund}},
	\bibinfo{author}{\bibfnamefont{W.~R.} \bibnamefont{Nelson}},
	\bibinfo{author}{\bibfnamefont{A.}~\bibnamefont{Abashian}},
	\bibinfo{author}{\bibfnamefont{C.}~\bibnamefont{Church}},
	\bibinfo{author}{\bibfnamefont{B.}~\bibnamefont{Lu}},
	\bibinfo{author}{\bibfnamefont{L.~W.} \bibnamefont{Mo}},
	\bibinfo{author}{\bibfnamefont{T.~A.} \bibnamefont{Nunamaker}},
	\bibnamefont{and} \bibinfo{author}{\bibfnamefont{P.}~\bibnamefont{Rassmann}},
	\bibinfo{journal}{Phys. Rev.} \textbf{\bibinfo{volume}{D38}},
	\bibinfo{pages}{3375} (\bibinfo{year}{1988}).
	
	\bibitem[{\citenamefont{De~Santis}(2015)}]{DeSantis:2015cma}
	\bibinfo{author}{\bibfnamefont{A.}~\bibnamefont{De~Santis}}
	(\bibinfo{collaboration}{KLOE-2, DAFNE Team}), \bibinfo{journal}{Phys.
		Scripta} \textbf{\bibinfo{volume}{T166}}, \bibinfo{pages}{014015}
	(\bibinfo{year}{2015}), \eprint{1503.06002}.
	
	\bibitem[{\citenamefont{He et~al.}(1991{\natexlab{a}})\citenamefont{He, Joshi,
			Lew, and Volkas}}]{He:1990pn}
	\bibinfo{author}{\bibfnamefont{X.~G.} \bibnamefont{He}},
	\bibinfo{author}{\bibfnamefont{G.~C.} \bibnamefont{Joshi}},
	\bibinfo{author}{\bibfnamefont{H.}~\bibnamefont{Lew}}, \bibnamefont{and}
	\bibinfo{author}{\bibfnamefont{R.~R.} \bibnamefont{Volkas}},
	\bibinfo{journal}{Phys. Rev.} \textbf{\bibinfo{volume}{D43}},
	\bibinfo{pages}{22} (\bibinfo{year}{1991}{\natexlab{a}}).
	
	\bibitem[{\citenamefont{He et~al.}(1991{\natexlab{b}})\citenamefont{He, Joshi,
			Lew, and Volkas}}]{He:1991qd}
	\bibinfo{author}{\bibfnamefont{X.-G.} \bibnamefont{He}},
	\bibinfo{author}{\bibfnamefont{G.~C.} \bibnamefont{Joshi}},
	\bibinfo{author}{\bibfnamefont{H.}~\bibnamefont{Lew}}, \bibnamefont{and}
	\bibinfo{author}{\bibfnamefont{R.~R.} \bibnamefont{Volkas}},
	\bibinfo{journal}{Phys. Rev.} \textbf{\bibinfo{volume}{D44}},
	\bibinfo{pages}{2118} (\bibinfo{year}{1991}{\natexlab{b}}).
	
	\bibitem[{\citenamefont{Kamenik and Smith}(2012)}]{Kamenik:2011vy}
	\bibinfo{author}{\bibfnamefont{J.~F.} \bibnamefont{Kamenik}} \bibnamefont{and}
	\bibinfo{author}{\bibfnamefont{C.}~\bibnamefont{Smith}},
	\bibinfo{journal}{JHEP} \textbf{\bibinfo{volume}{03}}, \bibinfo{pages}{090}
	(\bibinfo{year}{2012}), \eprint{1111.6402}.
	
	\bibitem[{\citenamefont{Davoudiasl
			et~al.}(2012{\natexlab{b}})\citenamefont{Davoudiasl, Lee, and
			Marciano}}]{Davoudiasl:2012ig}
	\bibinfo{author}{\bibfnamefont{H.}~\bibnamefont{Davoudiasl}},
	\bibinfo{author}{\bibfnamefont{H.-S.} \bibnamefont{Lee}}, \bibnamefont{and}
	\bibinfo{author}{\bibfnamefont{W.~J.} \bibnamefont{Marciano}},
	\bibinfo{journal}{Phys. Rev.} \textbf{\bibinfo{volume}{D86}},
	\bibinfo{pages}{095009} (\bibinfo{year}{2012}{\natexlab{b}}),
	\eprint{1208.2973}.
	
	\bibitem[{\citenamefont{Gardner et~al.}(2016)\citenamefont{Gardner, Holt, and
			Tadepalli}}]{Gardner:2015wea}
	\bibinfo{author}{\bibfnamefont{S.}~\bibnamefont{Gardner}},
	\bibinfo{author}{\bibfnamefont{R.~J.} \bibnamefont{Holt}}, \bibnamefont{and}
	\bibinfo{author}{\bibfnamefont{A.~S.} \bibnamefont{Tadepalli}},
	\bibinfo{journal}{Phys. Rev.} \textbf{\bibinfo{volume}{D93}},
	\bibinfo{pages}{115015} (\bibinfo{year}{2016}), \eprint{1509.00050}.
	
	\bibitem[{\citenamefont{Masjuan}(2012)}]{Masjuan:2012wy}
	\bibinfo{author}{\bibfnamefont{P.}~\bibnamefont{Masjuan}},
	\bibinfo{journal}{Phys. Rev.} \textbf{\bibinfo{volume}{D86}},
	\bibinfo{pages}{094021} (\bibinfo{year}{2012}), \eprint{1206.2549}.
	
	\bibitem[{\citenamefont{Escribano et~al.}(2015)\citenamefont{Escribano,
			Masjuan, and Sanchez-Puertas}}]{Escribano:2015nra}
	\bibinfo{author}{\bibfnamefont{R.}~\bibnamefont{Escribano}},
	\bibinfo{author}{\bibfnamefont{P.}~\bibnamefont{Masjuan}}, \bibnamefont{and}
	\bibinfo{author}{\bibfnamefont{P.}~\bibnamefont{Sanchez-Puertas}},
	\bibinfo{journal}{Eur. Phys. J.} \textbf{\bibinfo{volume}{C75}},
	\bibinfo{pages}{414} (\bibinfo{year}{2015}), \eprint{1504.07742}.
	
	\bibitem[{\citenamefont{Riordan et~al.}(1987)}]{Riordan:1987aw}
	\bibinfo{author}{\bibfnamefont{E.~M.} \bibnamefont{Riordan}}
	\bibnamefont{et~al.}, \bibinfo{journal}{Phys. Rev. Lett.}
	\textbf{\bibinfo{volume}{59}}, \bibinfo{pages}{755} (\bibinfo{year}{1987}).
	
	\bibitem[{\citenamefont{Gninenko}(2012{\natexlab{a}})}]{Gninenko:2011uv}
	\bibinfo{author}{\bibfnamefont{S.~N.} \bibnamefont{Gninenko}},
	\bibinfo{journal}{Phys. Rev.} \textbf{\bibinfo{volume}{D85}},
	\bibinfo{pages}{055027} (\bibinfo{year}{2012}{\natexlab{a}}),
	\eprint{1112.5438}.
	
	\bibitem[{\citenamefont{Boyce}(2012)}]{Boyce:2012ym}
	\bibinfo{author}{\bibfnamefont{J.~R.} \bibnamefont{Boyce}}
	(\bibinfo{collaboration}{HPS, DarkLight, LIPSS, APEX}), \bibinfo{journal}{J.
		Phys. Conf. Ser.} \textbf{\bibinfo{volume}{384}}, \bibinfo{pages}{012008}
	(\bibinfo{year}{2012}).
	
	\bibitem[{\citenamefont{Freytsis et~al.}(2010)\citenamefont{Freytsis,
			Ovanesyan, and Thaler}}]{Freytsis:2009bh}
	\bibinfo{author}{\bibfnamefont{M.}~\bibnamefont{Freytsis}},
	\bibinfo{author}{\bibfnamefont{G.}~\bibnamefont{Ovanesyan}},
	\bibnamefont{and} \bibinfo{author}{\bibfnamefont{J.}~\bibnamefont{Thaler}},
	\bibinfo{journal}{JHEP} \textbf{\bibinfo{volume}{01}}, \bibinfo{pages}{111}
	(\bibinfo{year}{2010}), \eprint{0909.2862}.
	
	\bibitem[{\citenamefont{Ilten et~al.}(2015)\citenamefont{Ilten, Thaler,
			Williams, and Xue}}]{Ilten:2015hya}
	\bibinfo{author}{\bibfnamefont{P.}~\bibnamefont{Ilten}},
	\bibinfo{author}{\bibfnamefont{J.}~\bibnamefont{Thaler}},
	\bibinfo{author}{\bibfnamefont{M.}~\bibnamefont{Williams}}, \bibnamefont{and}
	\bibinfo{author}{\bibfnamefont{W.}~\bibnamefont{Xue}},
	\bibinfo{journal}{Phys. Rev.} \textbf{\bibinfo{volume}{D92}},
	\bibinfo{pages}{115017} (\bibinfo{year}{2015}), \eprint{1509.06765}.
	
	\bibitem[{\citenamefont{Gninenko}(2012{\natexlab{b}})}]{Gninenko:2012eq}
	\bibinfo{author}{\bibfnamefont{S.~N.} \bibnamefont{Gninenko}},
	\bibinfo{journal}{Phys. Lett.} \textbf{\bibinfo{volume}{B713}},
	\bibinfo{pages}{244} (\bibinfo{year}{2012}{\natexlab{b}}),
	\eprint{1204.3583}.
	
	\bibitem[{\citenamefont{Brice{\~n}o et~al.}(2016)\citenamefont{Brice{\~n}o,
			Dudek, Edwards, Shultz, Thomas, and Wilson}}]{Briceno:2016kkp}
	\bibinfo{author}{\bibfnamefont{R.~A.} \bibnamefont{Brice{\~n}o}},
	\bibinfo{author}{\bibfnamefont{J.~J.} \bibnamefont{Dudek}},
	\bibinfo{author}{\bibfnamefont{R.~G.} \bibnamefont{Edwards}},
	\bibinfo{author}{\bibfnamefont{C.~J.} \bibnamefont{Shultz}},
	\bibinfo{author}{\bibfnamefont{C.~E.} \bibnamefont{Thomas}},
	\bibnamefont{and} \bibinfo{author}{\bibfnamefont{D.~J.}
		\bibnamefont{Wilson}}, \bibinfo{journal}{Phys. Rev.}
	\textbf{\bibinfo{volume}{D93}}, \bibinfo{pages}{114508}
	(\bibinfo{year}{2016}), \eprint{1604.03530}.
	
	\bibitem[{\citenamefont{Yamagata-Sekihara and
			Oset}(2010)}]{YamagataSekihara:2010ip}
	\bibinfo{author}{\bibfnamefont{J.}~\bibnamefont{Yamagata-Sekihara}}
	\bibnamefont{and} \bibinfo{author}{\bibfnamefont{E.}~\bibnamefont{Oset}},
	\bibinfo{journal}{Phys. Lett.} \textbf{\bibinfo{volume}{B690}},
	\bibinfo{pages}{376} (\bibinfo{year}{2010}), \eprint{1001.1816}.
	
\end{thebibliography}
\end{document}